\documentclass[11pt]{article}
\usepackage[numbers,sort&compress]{natbib} 
\usepackage[left=1.5cm, right=1.5cm, top=1.785cm, bottom=2.0cm]{geometry}
\usepackage{graphicx} 
\usepackage{caption}
\usepackage[usenames,dvipsnames]{xcolor}
\usepackage{amsmath}
\usepackage{hyperref}
\usepackage[noabbrev]{cleveref}
\usepackage{tikz}
\usepackage{subcaption}
\usepackage{siunitx}

\definecolor{outline}{RGB}{37,90,150}
\definecolor{gel}{RGB}{208,215,237}
\definecolor{cream}{RGB}{222,217,201}

\title{Will it bend? Modelling hydrogel box hinges with a gel Lattice Spring Method}
\author{Ellen M. Jolley\textsuperscript{*a} \and Katie A. Phillips\textsuperscript{a} \and Guy Harrop\textsuperscript{b} \and Gerard Cummins\textsuperscript{b,c} \and Tom D. Montenegro-Johnson\textsuperscript{a} \vspace{1cm} \\
\small
\textsuperscript{a} Mathematics Institute, Zeeman Building, University of Warwick, Coventry, CV4 7AL, UK \\
\small \textsuperscript{b} School of Engineering, University of Birmingham, Edgbaston, Birmingham, B15 2TT, UK\\
\small \textsuperscript{c} Healthcare Technologies Institute, University of Birmingham, Edgbaston, Birmingham, B15 2TT, UK
}

\begin{document}

\maketitle

\begin{abstract}
Hydrogels are biocompatible smart materials comprised of a hydrophilic polymer network which can absorb a large volume of water (swelling). The changes in shape the gel undergoes as it swells or de-swells can be utilised in the design of actuators. This is of particular interest for many applications whereby hydrogel actuators would be combined into a complex origami-inspired structure which can `fold' and `unfold' according to application needs. In this paper, we therefore focus on one such actuator design, the bilayer hinge, which can induce bending (for potential use in a `fold'/`unfold' mechanism). We develop a model using the gel lattice spring method (gLSM) which can be used to simulate composite structures comprised of hydrogel and elastomer and compute the equilibrium bending angle of bilayer hinges for a range of geometries and material properties.
\end{abstract}

\section{Introduction}

Transforming a compact, low profile configuration into a larger, functional geometry is a recurring engineering requirement wherever access, transport or storage volume is constrained. This requirement has underpinned the design of deployable structures for space applications \citep{wang_deployment_2024}, soft robotics \citep{mungekar_design_2023} and medical devices\citep{kuribayashi2006Selfdeployable,zhou2026Deployable}. Medicine presents a particularly challenging instance of this requirement as many procedures, including the  insertions of stents and catheters, biopsies, and surgeries, are highly invasive. As well as being uncomfortable, invasive procedures increase the risk of infection, slow recovery time and can cause tissue damage. As a result, there is continued interest in developing alternatives which reduce invasiveness \citep{klonowski2013Microincision,fuchs2002Minimally,fu2009Steerable}. An approach that has been continually investigated over the last decades is the use of deployable devices \citep{hu2018Soft}. Deployable medical devices (DMDs) are those which have some shape changing properties which enable them to be inserted in a compact form and then transform into the desired functional form once inside the body\citep{sokolowski2007medical,zhang2021Deployable,taylor2016Intracardiac}.

The advancement of 3D printing technology has accelerated the design of these devices, leading to an approach known as 4D printing \citep{tibbits20144d}. This combines 3D printing with the use of smart materials, such as shape-memory alloys, shape-memory polymers \citep{sokolowski2007medical,Scalet2020Two} or hydrogels that can deterministically swell or shrink in response to environmental factors such as temperature, light or pH \citep{dong2021Programmable,jiao2022Programmable} facilitating the design of actuators or sensors that change shape at any time in response to these environmental triggers\citep{mahmood2023revolutionizing}.
The simplest and most widely used actuator configuration is the bilayer in which a layer that swells is bonded to one that does not, and the resulting strain mismatch across the bond line is accommodated by curvature \citep{yang2015Hydrogel,yuk2017Hydraulic,maeda2007SelfWalking,zheng2018Mimosa,duan2017Bilayer,cheng2021recent,naficy20174d,bassik2010photolithographically,yoon2014functional,ahmed2020recent}.

The bilayer predates hydrogels, with bimetallic strips being the subject of Timoshenko's analysis \citep{timoshenko1925analysis}, which gives the curvature of a bonded pair of linearly elastic layers subjected to a uniform mismatch strain $\epsilon_{0}$ as

\begin{equation}
    \kappa = \frac{6\epsilon_0(1+m)^2}
    {h\left[3(1+m)^2 + (1+mn)\left(m^2 + \tfrac{1}{mn}\right)\right]}
    \label{eq:timoshenko}
\end{equation}

where $h$ is the total thickness, $m$ the ratio of layer thicknesses and $n$ the ratio of their Young's moduli; for equal thicknesses and moduli this reduces to $\kappa =3 \epsilon_{0}/2 h$. This equation establishes that the curvature is determined by the mismatch strain together with the thickness and elastic modulus ratio of the layers, and scales inversely with total thickness. It also highlights that the curvature of the bilayer is independent of the length of the layers, with the accumulated bend angle of uniformly curved strip given as $\theta = \kappa L$. These same equations apply equally to the swelling driven bending observed with hydrogel bilayers, with thermal mismatch replaced by the linear swelling strain. Hydrogel bilayers built on this principle have produced grippers, walkers and folding sheets\citep{bassik2010photolithographically, yoon2014functional, yang2015Hydrogel}, and the design rules that thinner layers and larger swelling ratios lead to greater curvature, while longer strips are needed for greater angles have been broadly confirmed experimentally \citep{naficy20174d}.

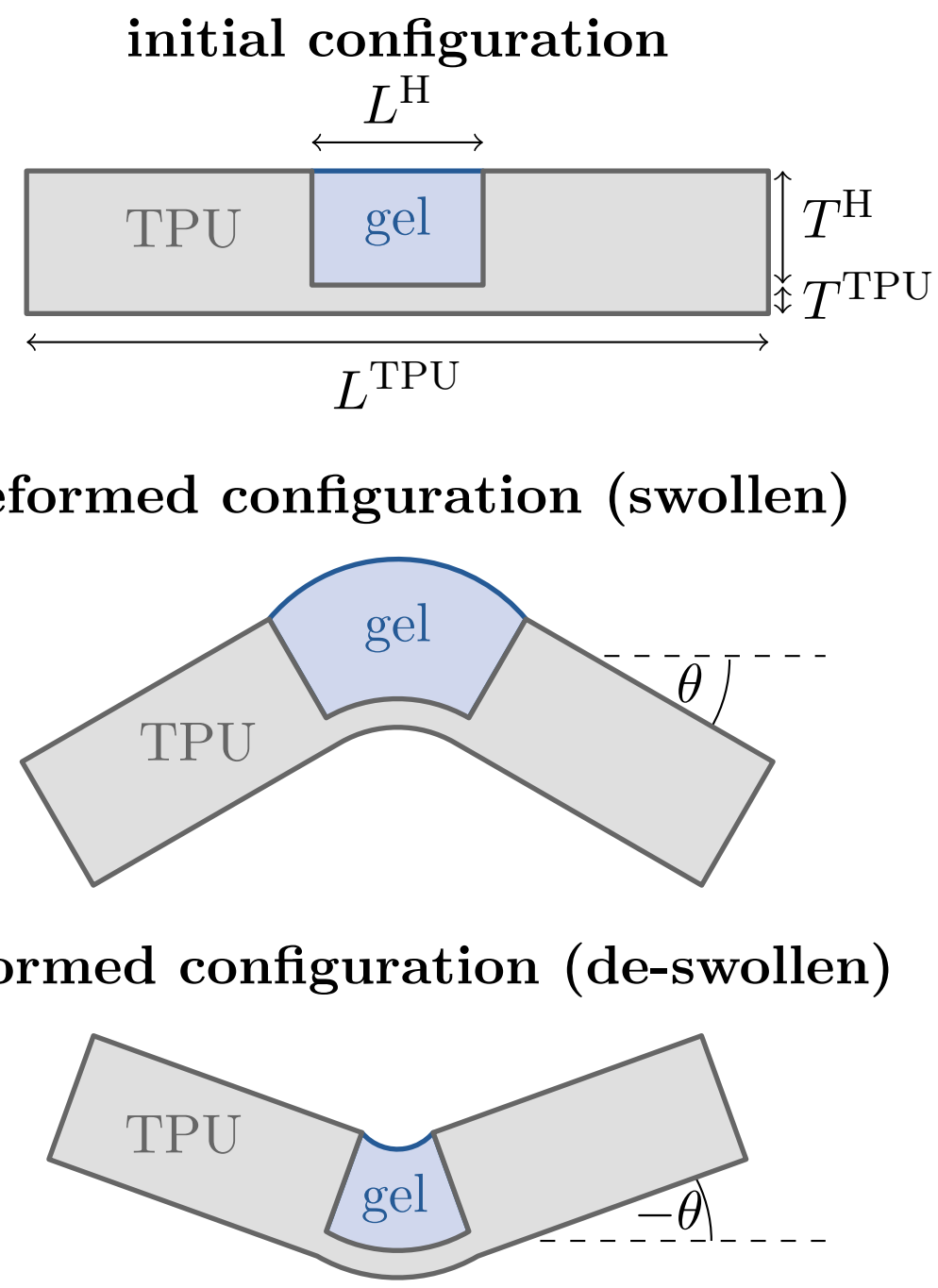
\begin{figure}
  \centering
  \begin{tikzpicture}
        %Initial Configuration
        \begin{scope}[shift={(-1,4)}]
            \fill[gel] (-0.6,1) -- (-0.6,1.8) -- (0.6,1.8) -- (0.6,1) -- cycle;
            \draw[outline,line width=1pt,line cap=round,line join=round]        (-0.6,1.8) 
                --(0.6,1.8);
            \node[outline] at (0,1.45) {gel};
            \draw[line width=1pt,line cap=round,line join=round,darkgray!80,fill=lightgray!50]
                (-2.6,1.8) 
                -- (-0.6,1.8) 
                -- (-0.6,1) 
                -- (0.6,1) 
                -- (0.6,1.8) 
                -- (2.6,1.8) 
                -- (2.6,0.8) 
                -- (-2.6,0.8) 
                -- cycle;
            \draw[<->] (-0.6,2) -- (0.6,2);
            \node[anchor=south] at (0,2) {$L^\text{H}$};
            \draw[<->] (-2.6,0.6) -- (2.6,0.6);
            \node[anchor=north] at (0,0.6) {$L^\text{TPU}$};
            \draw[<->] (2.7,1.01) -- (2.7,1.8);
            \draw[<->] (2.7,0.8) -- (2.7,0.99);
            \node[anchor=west] at (2.7,1.5) {$T^\text{H}$};
            \node[anchor=west] at (2.7,0.93) {$T^\text{TPU}$};
            \node[darkgray!80] at (-1.5,1.4) {TPU}; 
            \node at (0,2.7) {\textbf{initial configuration}};
            \node at (4,2.75) {};
        \end{scope}
        % Swollen Configuration
        \begin{scope}[shift={(-1,1.1)}]     
            \draw (2.2,0.8) arc (-30:0:0.95);
            \node at (2.05,1.11) {$\theta$};
            \draw[dashed] (1.45,1.3) -- (3,1.3);
            \fill[gel]
                ({-sin(30)},{cos(30)})
                -- ({-1.8*sin(30)},{1.8*cos(30)})
                arc (140:40:{1.8*sin(30)/sin(50)})
                -- ({sin(30)},{cos(30)})
                arc (60:120:1);
            \draw[outline,line width=1pt,line cap=round,line join=round]
                ({-sin(30)},{cos(30)})
                -- ({-1.8*sin(30)},{1.8*cos(30)})
                arc (140:40:{1.8*sin(30)/sin(50)})
                -- ({sin(30)},{cos(30)});
            \draw[line width=1pt,line cap=round,line join=round,darkgray!80,fill=lightgray!50]
                ({-0.8*sin(30)},{0.8*cos(30)})
                arc (120:60:0.8)
                -- +({2*cos(30)},{-2*sin(30)})
                -- +({2*cos(30)+sin(30)},{-2*sin(30)+cos(30)})
                -- +({sin(30)},{cos(30)})
                -- +({0.2*sin(30)},{0.2*cos(30)})
                arc (60:120:1)
                -- +({-0.8*sin(30)},{0.8*cos(30)})
                -- +({-2*cos(30)-0.8*sin(30)},{-2*sin(30)+0.8*cos(30)})
                -- +({-2*cos(30)+0.2*sin(30)},{-2*sin(30)-0.2*cos(30)})
                -- cycle;
            \node[outline] at (0,1.5) {gel};
            \node[darkgray!80] at (-1.4,0.7) {TPU};
            \node at (0,2.4) {\textbf{deformed configuration (swollen)}};
        \end{scope}
        % Deswollen Config
        \begin{scope}[shift={(-1,-2.5)}]
            \draw (2.2,0.8) arc (0:30:0.95);
            \node at (1.9,1) {$-\theta$};
            \draw[dashed] (1,0.8) -- (3,0.8);   
            \fill[gel]
                      ({-sin(30)},{cos(30)})
                      -- ({-0.5*sin(30)},{1.8*cos(30)})
                      arc (-140:-40:{0.5*sin(30)/sin(50)})
                      -- ({sin(30)},{cos(30)})
                      arc (-60:-120:1);
            \draw[outline,line width=1pt,line cap=round,line join=round]
                ({-sin(30)},{cos(30)})
                -- ({-0.5*sin(30)},{1.8*cos(30)})
                arc (-140:-40:{0.5*sin(30)/sin(50)})
                -- ({sin(30)},{cos(30)});
            \draw[line width=1pt,line cap=round,line join=round,darkgray!80,fill=lightgray!50]
                ({-1.125*sin(30)},{0.8*cos(30)})
                arc (-120:-60:1.125)
                -- ({1.125*sin(30)+2.172*cos(30)}, {0.8*cos(30)+1.358*sin(30)})
                -- ({0.5*sin(30)+2.172*cos(30)}, {1.8*cos(30)+1.358*sin(30)})
                -- ({0.5*sin(30)},{1.8*cos(30)})
                -- ({sin(30)},{cos(30)})
                arc (-60:-120:1)
                -- ({-0.5*sin(30)},{1.8*cos(30)})
                -- ({-0.5*sin(30)-2.172*cos(30)}, {1.8*cos(30)+1.358*sin(30)})
                -- ({-1.125*sin(30)-2.172*cos(30)}, {0.8*cos(30)+1.358*sin(30)})
                -- cycle;
            \node[outline] at (-0.02,1.1) {gel};
            \node[darkgray!80] at (-1.5,1.55) {TPU};
            \node at (0,2.7) {\textbf{deformed configuration (de-swollen)}};
        \end{scope}
    \end{tikzpicture}
    \caption{A schematic of a hydrogel box hinge. The hinge constructed of a cuboidal segment of hydrogel adhered to a surrounding TPU frame. The hinge begins in the undeformed configuration (top). Conditions are then changed and it reaches an equilibrium state and will either swell (middle) or de-swell (bottom) causing hinging with bend angle $\theta$.}
    \label{fig:hingediagram}
\end{figure}

The structures utilised here, shown in Figure \ref{fig:hingediagram}, depart from the classical bilayer. Instead of a uniformly curved strip, this box hinge structure consists of a compliant ligament connecting two comparatively rigid arms, with a block of hydrogel located in the resulting recess and adhered to the frame on three sides. This structure concentrates bending at the ligament rather than it being distributed along the strip, resulting in a discrete fold angle rather than a curvature. Localised bending such as this have been used in other self-folding microstructures \citep{bassik2010photolithographically, yoon2014functional}. Additionally, in a bimetallic strip the mismatch strain is determined by the two expansion coefficients and the temperature change, and does not depend on how stiff either layer is. A gel behaves differently as the stiffness of the polymer network decides how much it swells and hence how strongly the hinge is driven to bend and how much it resists bending. Designing these hydrogel box hinges therefore requires a precise understanding of how the geometry and mechanical properties of the hinge influence bending \citep{cheng2021recent,naficy20174d}. 

Mathematical models allow this parameter space to be explored efficiently and inexpensively. The standard in hydrogel modelling is based on Flory-Rehner theory \citep{flory1943statisticalA,flory1943statisticalB}, which conceptualises the gel as a two-phase material whose swelling dynamics are governed by a competition between chemical and elastic forces. Within this framework, some models take the approach of a reduction to linear elasticity \citep{doi2009Gel,webber2023linear} while others retain the fully nonlinear models \citep{bertrand2016dynamics,hong2008theory,kang2010Variational}. For the simplest geometries, solutions can be found analytically or by finite differences \citep{doi2009Gel,webber2023linear,bertrand2016dynamics}, but commonly finite element methods (FEM) are used to solve the resulting set of nonlinear coupled partial differential equations \citep{kang2010Variational,hong2008theory,chester2011thermo,bouklas2015Nonlinear}. Previous theoretical analyses of hydrogel bilayers have followed this trend, falling broadly into categories of analytic and semi-analytic models\citep{lucantonio2014Swellinginduced,nardinocchi2016Finite,morimoto2015Temperatureresponsive,abdolahi2016Analytical} and full numerical simulations by FEM \citep{amiri2023Transient,drozdov2016Swellinginduced}.

An alternative approach was developed by \citet{yashin2007Theoretical}. Known as the gel lattice spring method (gLSM), it is based on lattice spring models of elasticity \citep{ostoja-starzewski2002Lattice}. A lattice spring model can be conceptualised as a mesoscale model which resembles the underlying atomic lattice of the material though it does not directly capture the same microstructural physics. The simplest lattice spring models for linear elastic materials are constructed of Hookean springs and these can be shown to be algebraically equivalent to simple finite element methods \citep{ashurst1976Microscopic}. While they are relatives of finite element methods in a sense, there are a few significant benefits of the lattice approach for modelling actuators. Firstly, the method naturally handles material boundaries and geometric nonlinearity, with no need to enforce boundary conditions explicitly. Secondly, the computation of internal forces in a lattice spring method is easily physically interpreted, making it easier to adapt designs according to simulation results. Finally, the method readily lends itself to the inclusion of additional effects: for example, one could include chemical responsivity as considered by \citet{yashin2007Theoretical}, and similarly it is easier to include models of fracture mechanics in lattice spring approaches than in FEM. Lattice spring approaches therefore hold a high potential for developing complex models of deformable structures.

In this paper, we adapt the gLSM approach developed by \citet{yashin2007Theoretical} to modelling hydrogel bilayer hinges, and use the resulting model to investigate the effect of material properties and hinge geometry on the resultant bend angle at equilibrium. We also construct hydrogel box hinges and demonstrate their ability to produce significant bending angles in practice. In \cref{sec:setup}, we present the problem outline and detail our experiments. \Cref{sec:method} introduces the theoretical methodology, outlining the physics underlying hydrogel swelling and the principles of the gLSM. \Cref{sec:results} then presents results of the simulations. We discuss the relevance of the results to the application in \cref{sec:conc}.

\section{Problem specification}
\label{sec:setup}

\subsection{Experiment}

\begin{figure}[!h]
\centering
\begin{minipage}{0.32\linewidth}
    \centering
    \includegraphics[width=\linewidth]{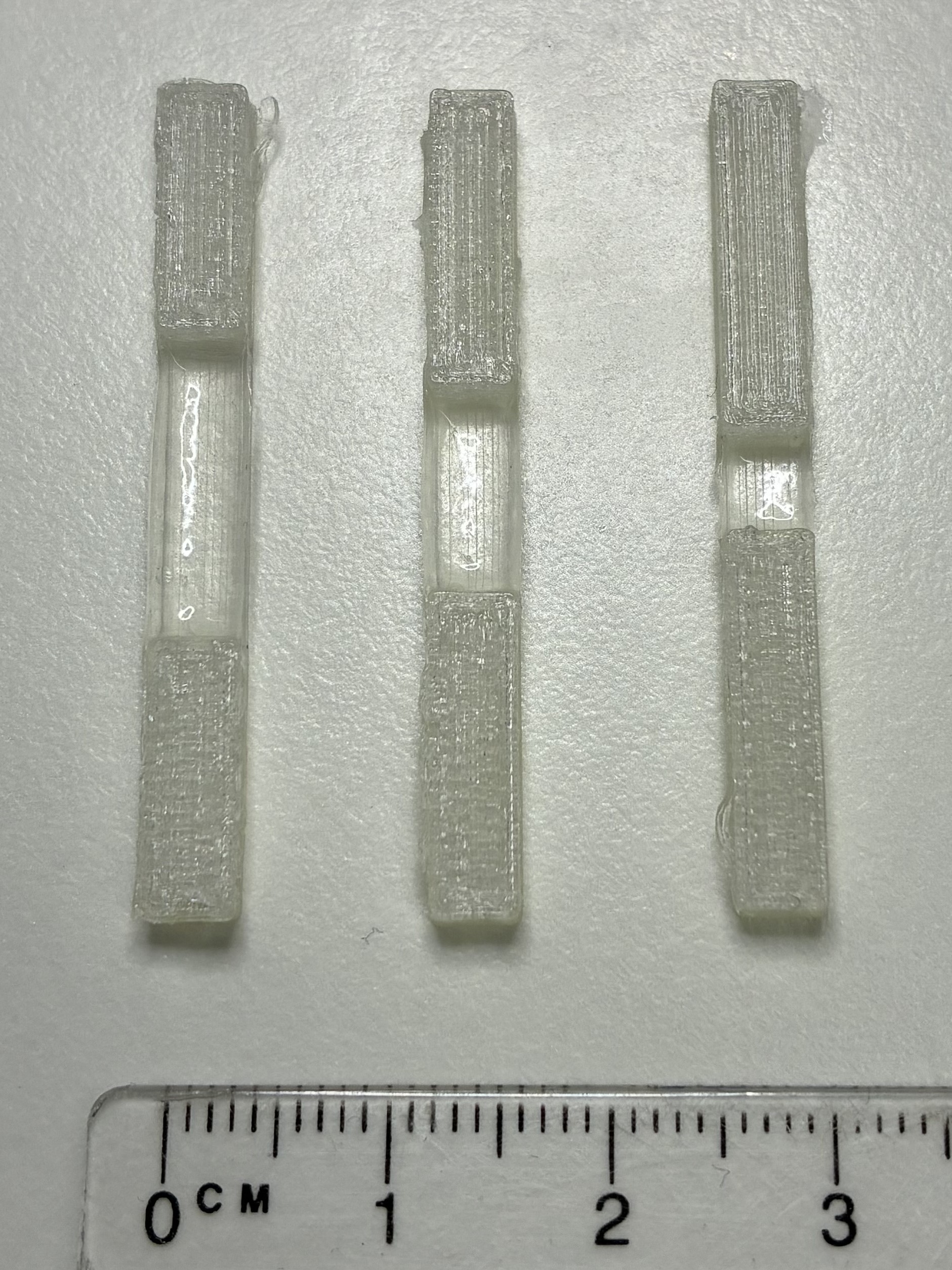}
\end{minipage}\hfill
\begin{minipage}{0.32\linewidth}
    \centering
    \includegraphics[width=\linewidth]{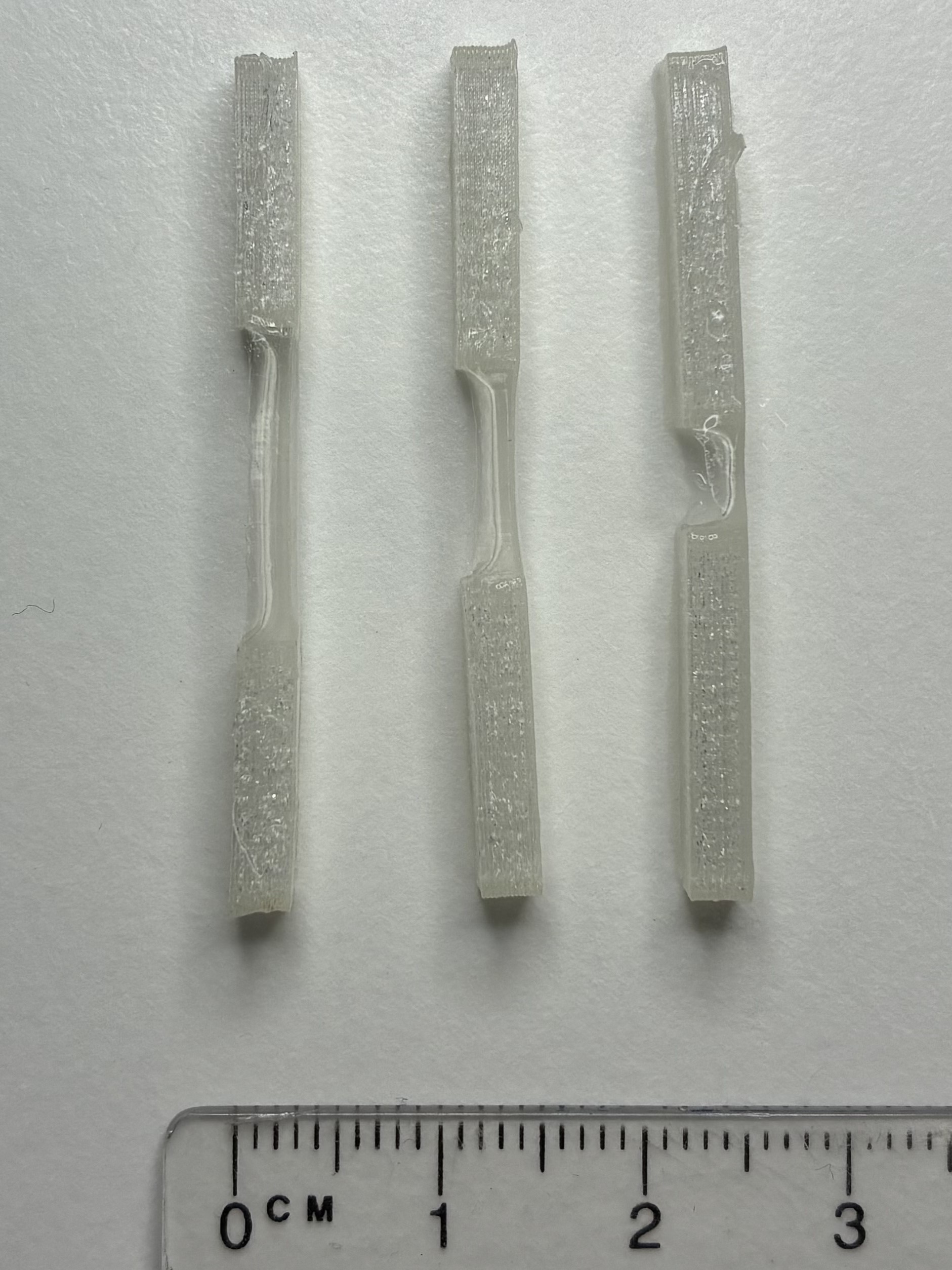}
\end{minipage}\hfill
\begin{minipage}{0.32\linewidth}
    \centering
    \includegraphics[width=\linewidth]{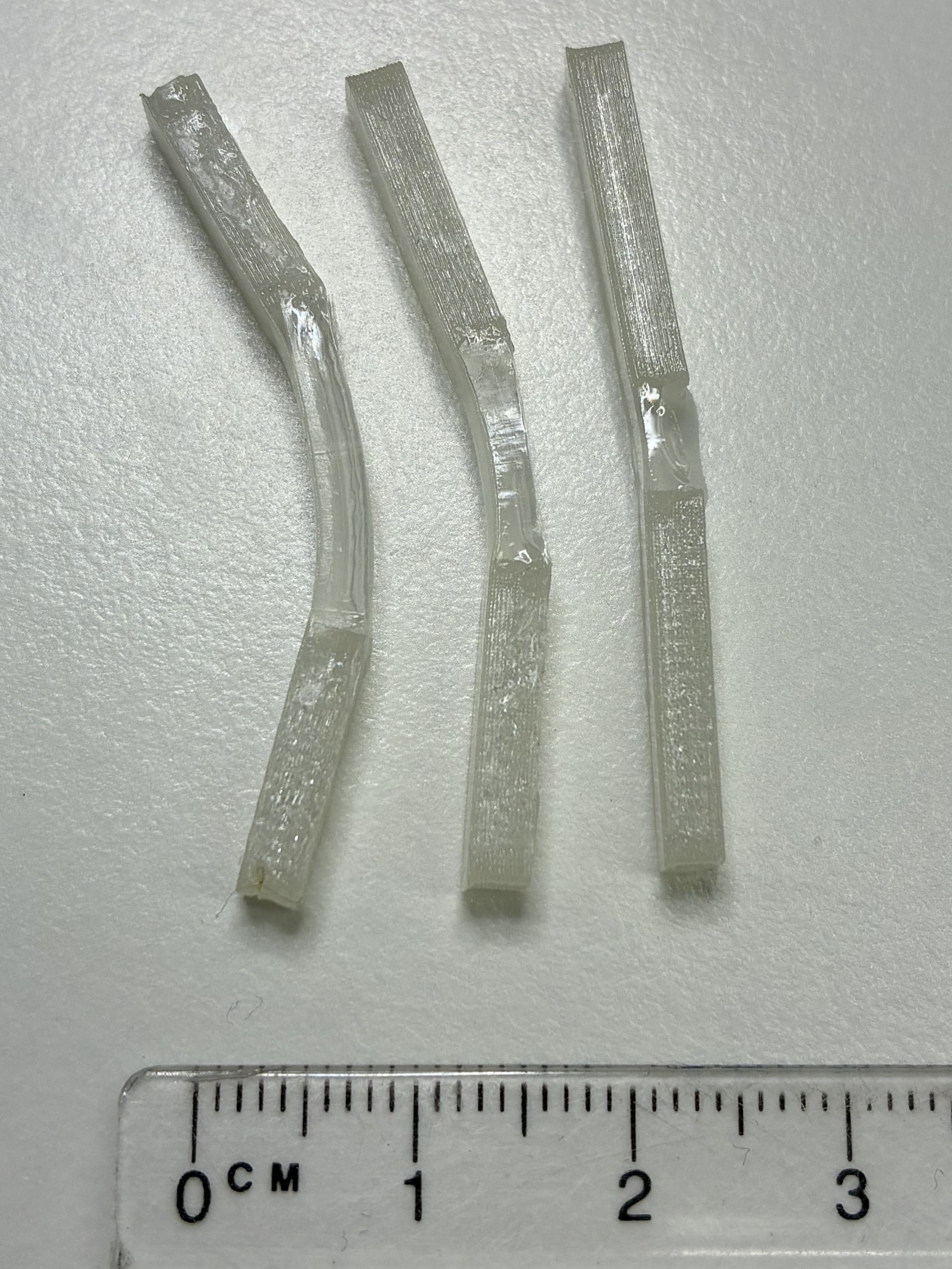}
\end{minipage}
\caption{Manufactured hydrogel box hinges with different hinge lengths. Top view of the dry hinges (left). Side view of the dry hinges (middle). Side view of the wet hinges (right).}
\label{fig:hinge_images_experimental}
\end{figure}

As proof of concept, we constructed hydrogel box hinges to assess their bending; the result is shown in \cref{fig:hinge_images_experimental}. An alginate hydrogel  strip (synthesis described below) is adhered on three sides to a frame composed of thermoplastic polyurethane (TPU; RS PRO TPU 98A, RS Components, Corby, UK). The swelling of the hydrogel upon submersion in water is then expected to cause the frame to bend. The TPU frames were printed on an Ender~3 fused filament fabrication (FFF) printer (Creality, Shenzhen, China) with a \SI{0.4}{\milli\metre} nozzle at \SI{225}{\celsius}, a layer thickness of \SI{0.2}{\milli\metre}, \SI{100}{\percent} infill and a print speed of \SI{20}{\milli\metre\per\second}. A borosilicate glass build plate provided sufficient adhesion without heating the bed. Printed parts were treated with an oxygen plasma (Zepto plasma cleaner, Diener Electronic, Ebhausen, Germany) at \SI{10}{\percent} power for \SI{10}{\minute} to modify the surface energy for improved adhesion with the hydrogel.

To synthesise the hydrogel, sodium alginate (Sigma-Aldrich, Gillingham, UK) was crosslinked by internal gelation using calcium carbonate (CaCO$_3$, Sigma-Aldrich) and glucono-$\delta$-lactone (GDL, food grade, MSK Ingredients, Chesterfield, UK). Sodium alginate (\SI{2}{\percent} w/v) was dissolved in deionised water over \SI{24}{\hour} under constant magnetic stirring at room temperature. CaCO$_3$ was added at \SI{20}{\percent} w$/$w and stirred for \SI{30}{\minute}. GDL was subsequently added to the solution and stirred for a further \SI{10}{\minute}. GDL hydrolyses slowly to gluconic acid, which dissolves the dispersed CaCO$_3$ and releases Ca$^{2+}$ uniformly throughout the solution, resulting in a homogeneous gel. A molar ratio of 2:1 (GDL:CaCO$_3$) was used to ensure complete dissolution of the CaCO$_3$ whilst keeping the pH close to neutral. The hinges were cast onto TPU frames with hinge lengths of \SI{5}{\milli\metre}, \SI{10}{\milli\metre}, and \SI{15}{\milli\metre} immediately after plasma treatment and allowed to crosslink for \SI{24}{\hour} in ambient conditions. The hinges were then imaged, submerged in deionised water for \SI{2}{\hour}, and imaged again (\cref{fig:hinge_images_experimental}). As can be seen in \cref{fig:hinge_images_experimental}, a longer strip of hydrogel increases bending.

\subsection{Problem set-up}
As a theoretical model of the box hinge design depicted in \cref{fig:hinge_images_experimental}, we a consider a two-dimensional vertical cross-section of the hinge and assume the deformation can be approximated by plane strain. The gel component is initially rectangular with length $L^\text{H}$ and thickness $T^\text{H}$. The surrounding TPU frame has thickness $T^\text{TPU}$ beneath the gel and $T^\text{H}+T^\text{TPU}$ at the arms. The total length of the hinge is $L^\text{TPU}$.

The swelling of the hydrogel is captured by the porosity $\phi$, a scalar field representing the local volume fraction of water compared to polymer at a given point. The hydrogel has initial porosity of $\phi = \phi_0$, which increases as the gel imbibes water. The effect of the swelling is to induce bending by two mechanisms: (1) the adhesion to the TPU below, causing mismatched strain associated with the bilayer effect, and (2) the upper exposed side of the gel is more inclined to swell than the lower, confined side. We denote the clockwise angle the arms make with the horizontal axis $\theta$. The goal of the current work is to compute the equilibrium configuration of the hinge and the resultant bend angle given the hinge geometry and material properties of the two components. We outline the methods by which we compute this angle and the theory of hydrogel swelling in \cref{sec:method} below.

\section{Methods}
\label{sec:method}
\subsection{Theory of hydrogel swelling}
The gel is conceptualised as a two-phase mixture, consisting of a solid polymer network phase and a fluid phase, each of which is taken to be individually incompressible. The porosity $\phi$ reflects the local volume fraction of fluid in the mixture. An Eulerian formulation is used, meaning quantities are described with respect to the current configuration of the gel at time $\tilde{t}$. 

We formulate our theoretical model of hydrogel dynamics using Flory-Rehner theory as a foundation. Flory-Rehner theory states the free energy density $\tilde{W}^\text{H}$ of the gel may be split in chemical and elastic components,
\begin{equation}
  \tilde{W}^\text{H} = \tilde{W}_{\text{mix}}+\tilde{W}_{\text{elast}}, \label{eq:WH}
\end{equation}
where $\tilde{W}_\text{mix}$ is the \textit{mixing free energy density} and $W_\text{elast}$ is the \textit{elastic} or \textit{strain energy density}. The dynamics of gel swelling and drying is then governed by the transfer of energy between these two components. Here and throughout the paper, $\tilde{\ }$ indicates a dimensional quantity while the absence thereof indicates a non-dimensional quantity.
\subsubsection{Elastic contribution}
The deformation of the gel is captured by a mapping $\tilde{\mathbf{X}} \to \tilde{\mathbf{x}}$ from the reference/undeformed configuration to the current/deformed configuration. We take the reference state to be the initial state of the gel (i.e. rectangular gel with porosity $\phi_0$). The elastic energy of the gel is given as a function of the deformation by the constitutive law, 
\begin{equation}
    \tilde{W}_{\text{elast}} = \frac{\tilde{\mu}^\text{H}_s}{2}\left[\mathrm{tr}\left(\mathbf{F}\mathbf{F}^T\right)-3-\log J\right], \label{eq:Welast}
\end{equation}
where $\tilde{\mu}_s$ is the shear modulus of the gel, $\mathbf{F} = \partial\mathbf{x}/\partial\mathbf{X}$ (in index notation, $F_{ij} = \partial x_i/\partial X_j$) is the deformation gradient tensor, and $J = \mathrm{det}\ \mathbf{F}$ is the Jacobian which represents the volumetric expansion ratio at a given point \citep{fung2017classical}.  We define the Terzaghi effective stress by
\begin{equation}
    \tilde{\boldsymbol\sigma}^\prime = J^{-1}\mathbf{F}\frac{\partial\tilde{W}_{\text{elast}}}{\partial\mathbf{F}} = \tilde{\mu}_sJ^{-1}\left(\mathbf{F}\mathbf{F}^T-\frac{1}{2}\mathbf{I}\right), \label{eq:effstress}
\end{equation}
which captures the contribution to the stress in gel due to strain the solid matrix. Note that this constitutive law differs from the standard neo-Hookean law in the presence of a factor of 2 in front of $\log J$ in \labelcref{eq:Welast}, which leads to the factor of $1/2$ in front of the identity in \labelcref{eq:effstress}. The effect of this is that in the reference state $\mathbf{F} = \mathbf{I}$, the gel network is experiencing elastic stress. Different hydrogel theories produce different coefficients from this term\citep{onuki1993Theory}; in the paper, we have followed \citet{yashin2007Theoretical} in taking this value. 
\subsubsection{Mixing contribution}
The mixing contribution to free energy is derived from Flory-Huggins polymer theory (see \citet{doi1995introduction}),
\begin{equation}
    \tilde{W}_{\text{mix}} = \tilde{K}J\left[\phi\log\phi+\chi\phi(1-\phi)\right],
\end{equation}
where the constant $\tilde{K} = \tilde{k}_B\tilde{T}/\tilde{\Omega}_f$ is the characteristic osmotic energy of the gel (where $\tilde{k}_B$ is the Boltzmann constant, $\tilde{T}$ is the temperature and $\tilde{\Omega}_f$ is the volume of fluid per fluid molecule), and $\chi$ is the Flory mixing parameter, which quantifies the interaction energy between polymer and solvent. We take the mixing parameter\citep{bertrand2016dynamics} to be a constant with the value 0.4. 

\subsubsection{Two-phase kinematics}
We assume both the water in the gel and the polymer scaffold are individually incompressible, and thus $J$ can be directly related to the porosity (which is equal to the local fluid volume fraction) by
\begin{equation}
  J = \frac{1-\phi_0}{1-\phi}, \label{eq:Jphi}
\end{equation}
where $\phi_0$ is the porosity of the initial configuration that is assumed constant, and $\phi$ is the porosity at each point in the current configuration. Mass conservation of the fluid and solid phases respectively dictate the balances,
\begin{equation}
    \frac{\partial\phi}{\partial \tilde{t}}+\tilde{\nabla}\cdot(\phi\tilde{\mathbf{v}}_f) = 0,
\label{eq:fluidcons} \end{equation} 
and
\begin{equation}
    \frac{\partial}{\partial \tilde{t}} (1-\phi)+\tilde{\nabla}\cdot[(1-\phi)\tilde{\mathbf{v}}_s] = 0, \label{eq:solidcons}
\end{equation}
where $\tilde{\mathbf{v}}_{f,s}$ are the fluid and solid velocities respectively. Together, \cref{eq:fluidcons,eq:solidcons} imply
  \begin{equation} \tilde{\nabla}\cdot\tilde{\mathbf{q}} = 0,\label{eq:flux}
  \end{equation}
where $\tilde{\mathbf{q}} = \phi\tilde{\mathbf{v}_f}+(1-\phi)\tilde{\mathbf{v}}_s$ represents the volume-averaged material flux. 

The interstitial fluid in the gel is assumed to be driven through the porous solid matrix by gradients in pore pressure $\tilde{p}$ according to Darcy's law,
\begin{equation}
    \phi(\tilde{\mathbf{v}}_f-\tilde{\mathbf{v}}_s) = -\frac{\tilde{k}(\phi)}{\tilde{\eta}_f}\tilde{\nabla}\tilde{p},\label{eq:darcy}
\end{equation}
where $\tilde{k}(\phi)$ is the permeability of the hydrogel, which depends on porosity, and $\tilde{\eta}_f$ is the dynamic viscosity of the fluid. We follow the hydrogel mechanics literature\citep{bertrand2016dynamics,yashin2007Theoretical} in adopting the following functional form for permeability,
\begin{equation}
  \tilde{k}(\phi) = \tilde{k}_0\phi(1-\phi)^{-3/2}.
\end{equation}
Combining \cref{eq:fluidcons,eq:solidcons,eq:flux,eq:darcy}, we arrive at a nonlinear advection--diffusion equation for the porosity of the gel,
\begin{equation}
    \frac{\partial\phi}{\partial \tilde{t}} +\tilde{\mathbf{q}}\cdot\tilde{\nabla}\phi= \tilde{\nabla}\cdot\left[\frac{(1-\phi)\tilde{k}(\phi)}{\tilde{\eta}_f}\tilde{\nabla}\tilde{p}\right]. \label{eq:adv-diff}
\end{equation}
\subsubsection{Momentum balance}
The total stress tensor in the gel is given by
\begin{align}
  \tilde{\boldsymbol\sigma}^\text{H} &= - \tilde{p}\mathbf{I} +\frac{\partial\tilde{W}_\text{mix}}{\partial J}\mathbf{I} +J^{-1}\mathbf{F}\frac{\partial\tilde{W}_\text{elast}}{\partial\mathbf{F}},\label{eq:stress} \\
  &= -(\tilde{p}+\tilde{\Pi})\mathbf{I}+\boldsymbol\sigma^\prime,
\end{align}
 where $\Pi$ is the \textit{osmotic pressure},
\begin{equation}
      \tilde{\Pi}(\phi) = -\frac{\partial\tilde{W}_{\text{mix}}}{\partial J} = -\tilde{K}\left[\log\phi+\chi(1-\phi)^2+(1-\phi)\right],
\end{equation}
and $\boldsymbol\sigma^\prime$ is the effective stress as given by \cref{eq:effstress}. The Cauchy momentum balance $\nabla\cdot\boldsymbol\sigma = \mathbf{0}$ then yields the partial differential equation (PDE),
\begin{equation}
      \tilde{\nabla} \tilde{p} +\nabla\tilde{\Pi}= \tilde{\nabla}\cdot\tilde{\boldsymbol\sigma}^\prime. \label{eq:cauchy}
\end{equation}
\subsubsection{Non-dimensionalization}
\label{sec:nond}
We non-dimensionalize our equations according to
\begin{gather}
    (\tilde{\mathbf{x}},\tilde{\boldsymbol\xi},\tilde{\mathbf{x}}_0) = \tilde{L}(\mathbf{x},\boldsymbol{\xi},\mathbf{x}_0), \quad (\tilde{\boldsymbol\sigma}^\text{H}, \tilde{p}, \tilde{\Pi}) = \tilde{K}(\boldsymbol\sigma^\text{H}, p, \Pi), \\ \tilde{k} = \tilde{k}_0k,\quad \tilde{t} = \frac{\tilde{L}^2\tilde{\eta}_f}{\tilde{k}_0\tilde{K}}t, \quad (\tilde{\mathbf{v}}_{f,s},\tilde{\mathbf{q}}) = \frac{\tilde{k}_0\tilde{K}}{\tilde{\eta}_f\tilde{L}}(\mathbf{v}_{f,s},\mathbf{q}), \notag
\end{gather}
where $\tilde{L}$ is some measure of the size of the hydrogel, and $\tilde{k}_0$ is the permeability scale. The composite $\tilde{L}^2\tilde{\eta}_f/\tilde{k}_0\tilde{K}$ is known as the poroelastic timescale.

The governing equations for the swelling of the gel are then given in non-dimensional form as follows. The momentum equation in the gel is 
\begin{equation}
    \nabla (p+\Pi)  = \nabla\cdot\boldsymbol\sigma', \label{eq:sys1}
\end{equation}
where $\Pi(\phi) = -[\log\phi+\chi(1-\phi)^2+(1-\phi)]$. This is coupled to the constitutive law, 
\begin{equation}\label{eq:sys2}
    \boldsymbol\sigma' = G^HJ^{-1}(\mathbf{FF}^T-\frac{1}{2}\mathbf{I}),
\end{equation}
where $G^H = \tilde{\mu}^\text{H}_s/\tilde{K} = \tilde{\Omega}_p/\tilde{\Omega_f}$ is a dimensionless number which captures the relative importance of elastic to chemical forces. The Jacobian and deformation gradient are related to displacement and porosity via
\begin{equation}
    \mathbf{F}^{-1} = \mathbf{I}-\nabla\boldsymbol\xi, \quad J = \det{\mathbf{F}} = \frac{1-\phi_0}{1-\phi}.\label{eq:sys3}
\end{equation}
This is coupled to the advection--diffusion equation for porosity,
\begin{equation}
    \frac{\partial{\phi}}{\partial t} +\mathbf{q}\cdot\nabla\phi = \nabla\cdot\left[(1-\phi)k(\phi)\nabla p\right],\label{eq:sys4}
\end{equation}
where
\begin{equation}
    \mathbf{q} = -k(\phi)\nabla p+\mathbf{v}_s, \quad \mathbf{v}_s = \frac{\partial\boldsymbol\xi}{\partial t}\cdot\mathbf{F}, \quad \nabla\cdot\mathbf{q} = 0.\label{eq:sys5}
\end{equation}
\subsection{TPU model}
We treat the TPU as a compressible neo-Hookean solid with strain energy density \citep{bonet2008Nonlinear}
\begin{equation}
  \tilde{W} = \frac{\tilde{\mu}^\text{TPU}_s}{2}\left[\mathrm{tr}\left(\mathbf{F}\mathbf{F}^T\right)-3-2\log J\right]+\frac{\tilde{\lambda}}{2}(\log J)^2, \label{eq:WTPU}
\end{equation}
where $\tilde{\lambda}$ is the bulk modulus of the TPU. The non-dimensional stress tensor is then given by
\begin{equation}
  \boldsymbol{\sigma}^\text{TPU} = G^\text{TPU}J^{-1}\left(\mathbf{F}\mathbf{F}^T-\mathbf{I}\right)+\lambda J^{-1}\log J,
\end{equation}
where $(G^\text{TPU},\lambda) = (\tilde{\mu}_s^\text{TPU},\tilde{\lambda})/\tilde{K}$. The momentum balance simply reads
\begin{equation}
  \nabla\cdot\boldsymbol\sigma^\text{TPU} = \mathbf{0}.
\end{equation}
\subsection{Lattice spring model}
A lattice spring model is a common computational technique in elasticity based on discretizing a material into a network of a springs \citep{ostoja-starzewski2002Lattice}.  We adapt an approach developed by \citet{yashin2007Theoretical} for chemo-responsive gels which we use in different forms for both the gel and the TPU; we outline the details for each in \cref{sec:gel,sec:TPU} below. We are using a 2D method in this work but the method can be adapted to three dimensions\citep{kuksenok2008Threedimensional}.
\begin{figure}
  \centering
  \begin{subfigure}{\linewidth}
    \centering
  \begin{tikzpicture}[scale=1.2, every dot/.style={    circle,
    fill,
    inner sep=1.5pt,
    node contents={}
  }]

\node[anchor=north east] at (0,0) {$\mathbf{x}_{(k-1,l-1)}$}; \filldraw [black] (0,0) circle (2pt);
\node[anchor=north] at (1,0) {$\mathbf{x}_{(k,l-1)}$}; \filldraw [black] (0,1) circle (2pt);
\node[anchor=north west] at (2,0) {$\mathbf{x}_{(k+1,l-1)}$}; \filldraw [black] (0,2) circle (2pt);
\node[anchor=east] at (0,1) {$\mathbf{x}_{(k-1,l)}$}; \filldraw [black] (1,0) circle (2pt);
\node at (1.5,1.11) {$\mathbf{x}_{(k,l)}$}; \filldraw [black] (1,1) circle (2pt);
\node[anchor=west] at (2,1) {$\mathbf{x}_{(k+1,l)}$}; \filldraw [black] (1,2) circle (2pt);
\node[anchor=south east] at (0,2) {$\mathbf{x}_{(k-1,l+1)}$}; \filldraw [black] (2,0) circle (2pt);
\node[anchor=south] at (1,2) {$\mathbf{x}_{(k,l+1)}$}; \filldraw [black] (2,1) circle (2pt);
\node[anchor= south west] at (2,2) {$\mathbf{x}_{(k+1,l+1)}$}; \filldraw [black] (2,2) circle (2pt);

\draw[semithick] (0,0) -- (1,1);
\draw (0,1) -- (1,1);
\draw (1,0) -- (1,1);
\draw (2,0) -- (1,1);
\draw (1,2) -- (1,1);
\draw (0,2) -- (1,1);
\draw (2,1) -- (1,1);
\draw (2,2) -- (1,1);

\draw[dashed,semithick] (0,0) -- (1,0);
\draw[dashed,semithick] (0,0) -- (0,1);
\draw[dashed,semithick] (1,0) -- (2,0);
\draw[dashed,semithick] (2,0) -- (2,1);
\draw[dashed,semithick] (0,2) -- (0,1);
\draw[dashed,semithick] (0,2) -- (1,2);
\draw[dashed,semithick] (2,2) -- (2,1);
\draw[dashed,semithick] (2,2) -- (1,2);

  \end{tikzpicture}
  \caption{A diagram showing spring-like contributions to the force on node $(k,l)$ due to its nearest neighbours.}
  \label{fig:spring}
\end{subfigure}
\begin{subfigure}{\linewidth}
  \centering
  \begin{tikzpicture}[scale=1.2, every dot/.style={    circle,
    fill,
    inner sep=1.5pt,
    node contents={}
  }]

\filldraw [black] (0,0) circle (2pt);
\filldraw [black] (0,1) circle (2pt);
\filldraw [black] (0,2) circle (2pt);
\filldraw [black] (1,0) circle (2pt);
\filldraw [black] (1,1) circle (2pt);
\filldraw [black] (1,2) circle (2pt);
\filldraw [black] (2,0) circle (2pt);
\filldraw [black] (2,1) circle (2pt);
\filldraw [black] (2,2) circle (2pt);

\draw (0,1) -- (1,1);
\draw (1,0) -- (1,1);
\draw (1,2) -- (1,1);
\draw (2,1) -- (1,1);

\draw[dashed,semithick] (0,0) -- (1,0);
\draw[dashed,semithick] (0,0) -- (0,1);
\draw[dashed,semithick] (1,0) -- (2,0);
\draw[dashed,semithick] (2,0) -- (2,1);
\draw[dashed,semithick] (0,2) -- (0,1);
\draw[dashed,semithick] (0,2) -- (1,2);
\draw[dashed,semithick] (2,2) -- (2,1);
\draw[dashed,semithick] (2,2) -- (1,2);

\node at (1.5,1.5) {$P_{(k,l)}$};
\node at (0.5,0.5) {\small $P_{(k-1,l-1)}$};
\node at (0.5,1.5) {$P_{(k-1,l)}$};
\node at (1.5,0.5) {$P_{(k,l-1)}$};

\draw[->] (1.35,1.35) -- (1.15,1.15);
\draw[->] (0.65,0.65) -- (0.85,0.85);
\draw[->] (0.65,1.35) -- (0.85,1.15);
\draw[->] (1.35,0.65) -- (1.15,0.85);
  \end{tikzpicture}
  \caption{A diagram showing the pressure-like contributions to the force on node $(k,l)$ due to the pressures from surrounding elements; elements are given labels equal to that of their bottom-left node.}
  \label{fig:pressure}
\end{subfigure}
\caption{Spring-like (a) and pressure-like (b) contributions to nodal forces.}
\label{fig:mesh}
\end{figure}
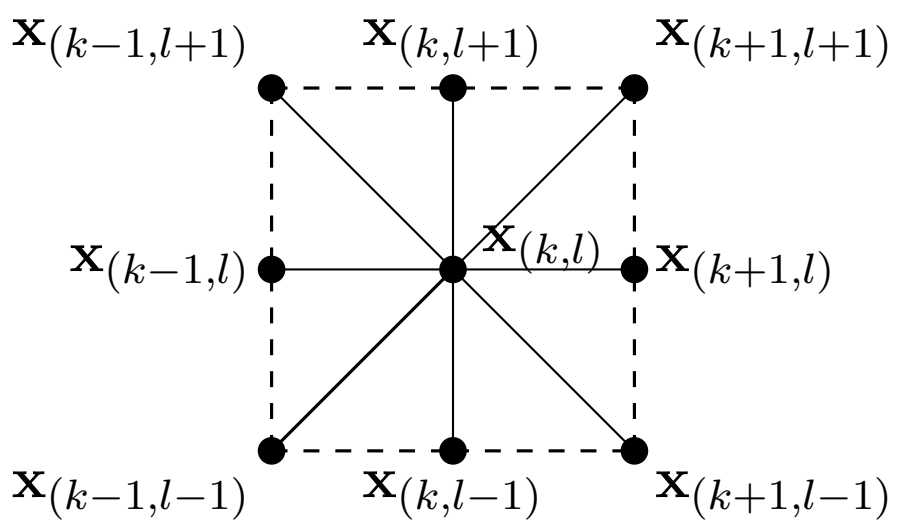
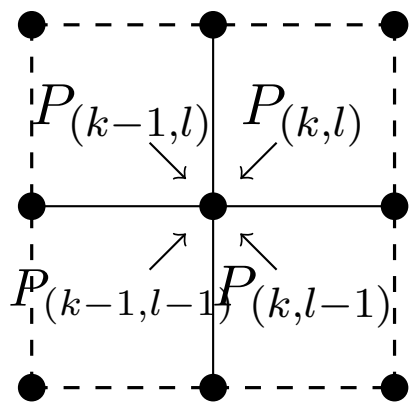

\subsubsection{Hydrogel}
\label{sec:gel}
For the hydrogel, our method is identical to the gel lattice spring model (gLSM) developed by \citet{yashin2007Theoretical} but with the chemical concentrations of both reagents set identically to zero to remove effects of chemical responsivity. The gel is discretized into a mesh of nodes with positions $\mathbf{x}_{(k,l)}$ and each node experiences a spring-like force as a result of interactions with its nearest neighbours (including diagonal neighbours) alongside pressure like contributions from neighbouring elements; see \cref{fig:mesh}. This form for nodal forces is derived by first expressing the total energy of the system as a function of nodal coordinates $U(\{\mathbf{x}_\mathbf{n}\})$ where $\mathbf{n} = (k,l)$ and then computing the nodal force directly as $\mathbf{F}_\mathbf{n} = -\partial U/\partial \mathbf{x}_\mathbf{n}$. The total free energy is obtained as a sum of elemental free energy densities,
\begin{equation}
  U = \sum_\mathbf{m} W_\text{el}(\mathbf{m}).
\end{equation}
Here $\mathbf{m} = (k,l)$ is the index of the element whose bottom-left node has index $\mathbf{n} = (k,l)$; we use $\mathbf{n}$ for nodes and $\mathbf{m}$ for elements to distinguish clearly between nodal and elemental quantities. The elemental energy density is obtained as
\begin{equation}
  W^\text{H}_\text{el}(\mathbf{m}) = \frac{1}{4}\int W^\text{H}\ \mathrm{d}\Omega_\mathbf{m}, \label{eq:WHel}
\end{equation}
where $\mathrm{d}\Omega_\mathbf{m}$ indicates an integral over the area of element $\mathbf{m}$ and $W^\text{H}$ is as given by \cref{eq:WH}. A local approximation of the deformation tensor via shape functions \citep{yashin2007Theoretical} eventually leads to the expression
\begin{multline}
  W_\text{el}(\mathbf{m}) = G^\text{H}\Delta^{-2}\sum_{1\leq n < n'\leq 4}\left[ K_{n,n'}(\mathbf{x}_n(\mathbf{m})-\mathbf{x}_n'(\mathbf{m}))^2\right]\\
  -\frac{1}{2}G^\text{H}\log J(\mathbf{m})+W_\text{mix}(\mathbf{m}),
\end{multline}
where $\mathbf{x}_n(\mathbf{m})$ for $n = 1,2,3,4$ are the coordinates of the four nodes composing element $\mathbf{m}$, $\Delta$ is the (constant) spacing between nodes in the reference configuration,  and 
\begin{equation}
  K_{n,n'} = \begin{cases}
    1/3 & \text{for edges,}\\
    1/6 & \text{for diagonals.} \label{eq:stiffness}
  \end{cases}
\end{equation}
The elemental Jacobian $J(\mathbf{m})$ can be calculated straightforwardly from the nodal coordinates of the element (it is the ratio of area in the current configuration to the reference configuration). This results in the forces
\begin{multline}
  \mathbf{F}_{(k,l)} = \frac{G^\text{H}}{3}\sum_{\delta,\delta'=0,\pm 1} \left[(\mathbf{x}_{(k+\delta,l+\delta')}-\mathbf{x}_{(k,l)})\right] \\+ \frac{1}{2}\mathbf{e}_3 \times \left[ (\mathbf{x}_{(k,l+1)}-\mathbf{x}_{(k+1,l)})P_{(k,l)}+(\mathbf{x}_{(k-1,l)}-\mathbf{x}_{(k,l+1)})P_{(k-1,l)}\right. \\ \left.+(\mathbf{x}_{(k,l-1)}-\mathbf{x}_{(k-1,l)})P_{(k-1,l-1)}+(\mathbf{x}_{(k+1,l)}-\mathbf{x}_{(k,l-1)})P_{(k,l-1)}\right]. \label{eq:nodalforce}
\end{multline}
The pressure of element $\mathbf{m}$ is given by
\begin{equation}
  P_\mathbf{m} = \Pi(\phi_\mathbf{m})+\frac{1}{2}G^\text{H}(1-\phi_\mathbf{m})/(1-\phi_0), \label{eq:gelpressure}
\end{equation}
so includes the effects of both osmotic pressure and isotropic elastic forces. \Cref{eq:nodalforce} therefore covers the effects of osmotic pressure $\Pi$ and effective elastic stress $\boldsymbol\sigma^\prime$ in \cref{eq:sys1}. 

For the remaining term in \labelcref{eq:sys1} , $\nabla p$, we use Darcy's law \labelcref{eq:darcy} to convert this to an expression of nodal velocity. In order to do this, we take the additional assumption that the net material flux $\mathbf{q}$ in the gel is zero and hence
\begin{equation}
  \phi\mathbf{v}_f+(1-\phi)\mathbf{v}_s = \mathbf{0}. \label{eq:qzero}
\end{equation}
For free swelling gels, it can commonly be demonstrated analytically that this must be the case from the equation $\nabla\cdot\mathbf{q} = 0$. For cases where the gel is adhered to or confined by rigid surfaces, it is typically not zero; for example, the solution derived by \citet{jolley2026Asymptotic} necessarily has non-zero $\mathbf{q}$ at the adhered boundary since the solid may not move but fluid still enters. However, since $\mathbf{q}$ is necessarily zero at equilibrium, we do not expect this to affect the final equilibrium states of the box hinges.

\Cref{eq:qzero,eq:darcy} together can be used to relate solid velocity to the pressure gradient which may then be used to relate solid velocity to elastic and osmotic forces by \cref{eq:sys1}. Integrating the result yields an expression for the velocity of each node in terms of the nodal force,
\begin{equation}
  \frac{\mathrm{d}\mathbf{x}_\mathbf{n}}{\mathrm{d}t} = M_\mathbf{n}\mathbf{F}_\mathbf{n}, \label{eq:timeint}
\end{equation}
where the nodal mobility coefficient $M^\text{H}_\mathbf{n}$ is given by
\begin{equation}
  M^\text{H}_\mathbf{n} = 4\Delta^{-2}k(\phi_\mathbf{n}) = 4\Delta^{-2}k_0\phi_\mathbf{n}(1-\phi_\mathbf{n})^{-3/2}.
\end{equation}
Here $\phi_\mathbf{n}$ is the nodal porosity, approximated from the elemental porosity as
\begin{equation}
  \phi_\mathbf{n} = \frac{1}{4}\sum_{\mathbf{m} \in E_\mathbf{n}} \phi_\mathbf{m},
\end{equation}
where $E_\mathbf{n}$ is the set of elements containing node $\mathbf{n}$. The elemental porosity $\phi_m$ is evaluated from the elemental Jacobian $J(\mathbf{m})$ via \cref{eq:Jphi}.
\label{sec:gLSM}
\subsubsection{TPU}
The free energy density of TPU \labelcref{eq:WTPU} has the same form as the elastic free energy density of the hydrogel \labelcref{eq:Welast} but with an added contribution to isotropic elastic stress which captures compressibility. We therefore adapt the form given for the elemental energy density in the hydrogel \labelcref{eq:WHel} for the TPU as 
\begin{multline}
  W^\text{TPU}_\text{el}(\mathbf{m}) = G^\text{TPU}\Delta^{-2}\sum_{1\leq n < n'\leq 4} K_{n,n'}(\mathbf{x}_n(\mathbf{m})-\mathbf{x}_n'(\mathbf{m}))^2\\
  -G^\text{TPU}\log J(\mathbf{m})+\frac{\lambda}{2} (\log J(\mathbf{m}))^2,
\end{multline}
which can be similarly intepreted as a combined spring-like contribution due to deviatoric stress and pressure-like contrbitution due to isotropic stress. This leads to a similar form for nodal forces as \cref{eq:nodalforce} but with a different stiffness parameter $G^\text{TPU}$ and an alternate form for pressure,
\begin{equation}
  P_\mathbf{m} = G^\text{TPU}J(\mathbf{m})^{-1}+\lambda J(\mathbf{m})^{-1}\log J(\mathbf{m}). \label{eq:tpupressure}
\end{equation}
We solve for the velocity of TPU nodes using the same form as \labelcref{eq:timeint},
\begin{equation}
  \frac{\mathrm{d}\mathbf{x}_\mathbf{n}}{\mathrm{d}t} = M^\text{TPU}_\mathbf{n} F_\mathbf{n}\label{eq:timeint2}
\end{equation}
In order to fully capture the dynamics of the box hinge, an appropriate assumption would be that the TPU is in effective equilibrium at each time-step since the diffusion-limited process of gel swelling is much slower than the solid equilibration time-scale \citep{tanaka1979kinetics}. This could be treated by separate convergence loops for each material on each time step. However, since we noted in \cref{sec:gLSM} that the assumption $\mathbf{q} = \mathbf{0}$ is likely inappropriate for physical gel dynamics in a box hinge, we focus on obtaining equilibrium states for the coupled system. Given this, we simply use the same time-stepping for both the TPU and the gel and interpret the time-coordinate as a purely fictitious numerical relaxation device \citep{otter1965Computations}.
\label{sec:TPU}
\subsection{Implementation}
\label{sec:implementation}
To summarise the method, we discretise the box hinge into an initially uniform, two-dimensional mesh composed of square elements, as shown in \cref{fig:mesh}. Each element is designated as either `TPU' or `gel' according to its initial location in the mesh. They are also assigned a pressure function, either \labelcref{eq:tpupressure} or \labelcref{eq:gelpressure} according to this designation. 

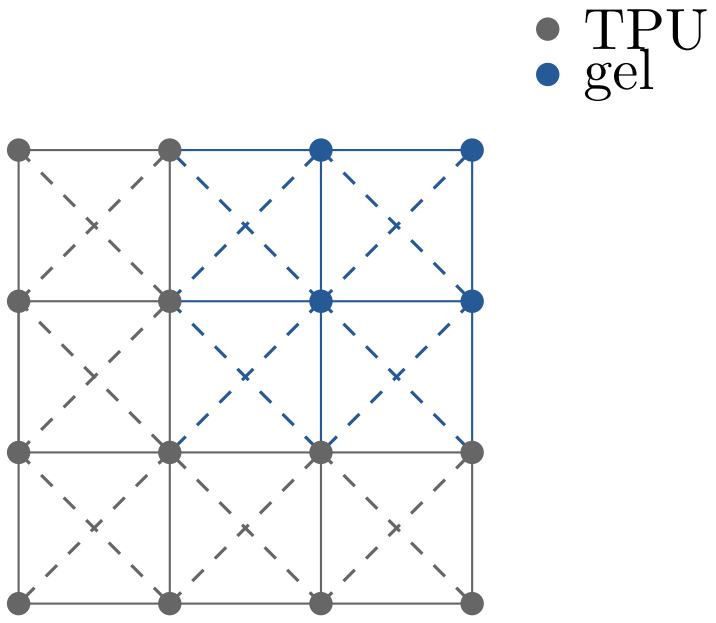
\begin{figure}
  \centering
  \begin{tikzpicture}

\draw[darkgray!80] (0,0) -- (1,0) -- (2,0) -- (3,0);
\draw[darkgray!80] (0,0) -- (0,1) -- (0,2) -- (0,3);
\draw[darkgray!80] (1,0) -- (1,1) -- (1,2) -- (1,3);
\draw[darkgray!80] (0,1) -- (1,1) -- (2,1) -- (3,1);
\draw[darkgray!80] (2,0) -- (2,1);
\draw[darkgray!80] (0,2) -- (0,1);
\draw[darkgray!80] (0,2) -- (1,2);
\draw[darkgray!80] (0,3) -- (1,3);
\draw[darkgray!80] (3,0) -- (3,1);

\draw[outline] (1,3) -- (2,3) -- (3,3);
\draw[outline] (1,2) -- (2,2) -- (3,2);
\draw[outline] (2,1) -- (2,2) -- (2,3);
\draw[outline] (3,1) -- (3,2) -- (3,3);

\draw[outline,dashed,semithick] (2,2) -- (1,1);
\draw[outline,dashed,semithick] (2,2) -- (3,3);
\draw[outline,dashed,semithick] (2,2) -- (1,3);
\draw[outline,dashed,semithick] (2,2) -- (3,1);
\draw[outline,dashed,semithick] (1,2) -- (2,1);
\draw[outline,dashed,semithick] (1,2) -- (2,3);
\draw[outline,dashed,semithick] (2,3) -- (3,2);
\draw[outline,dashed,semithick] (3,2) -- (2,1);

\draw[darkgray!80,dashed,semithick] (2,0) -- (1,1) -- (0,2);
\draw[darkgray!80,dashed,semithick] (1,1) -- (0,0);
\draw[darkgray!80,dashed,semithick] (0,1) -- (1,2);
\draw[darkgray!80,dashed,semithick] (0,2) -- (1,3);
\draw[darkgray!80,dashed,semithick] (1,0) -- (2,1);
\draw[darkgray!80,dashed,semithick] (2,0) -- (3,1);
\draw[darkgray!80,dashed,semithick] (2,1) -- (3,0);
\draw[darkgray!80,dashed,semithick] (0,3) -- (1,2);
\draw[darkgray!80,dashed,semithick] (0,1) -- (1,0);

\filldraw [darkgray!80] (0,0) circle (2pt);
\filldraw [darkgray!80] (0,1) circle (2pt);
\filldraw [darkgray!80] (0,2) circle (2pt);
\filldraw [darkgray!80] (0,3) circle (2pt);
\filldraw [darkgray!80] (1,0) circle (2pt);
\filldraw [darkgray!80] (1,1) circle (2pt);
\filldraw [darkgray!80] (1,2) circle (2pt);
\filldraw [darkgray!80] (1,3) circle (2pt);
\filldraw [darkgray!80] (2,0) circle (2pt);
\filldraw [darkgray!80] (2,1) circle (2pt);
\filldraw [outline] (2,2) circle (2pt);
\filldraw [outline] (2,3) circle (2pt);
\filldraw [darkgray!80] (3,0) circle (2pt);
\filldraw [darkgray!80] (3,1) circle (2pt);
\filldraw [outline] (3,2) circle (2pt);
\filldraw [outline] (3,3) circle (2pt);

\filldraw [darkgray!80] (3.5,3.8) circle (2pt);
\filldraw [outline] (3.5,3.5) circle (2pt);

\node[anchor=west] at (3.6,3.8) {TPU};
\node[anchor=west] at (3.6,3.5) {gel};

\node[anchor=east] at (-0.6,3.8) {};
  \end{tikzpicture}

  \caption{An illustration showing the treatment of nodes at the gel-TPU boundary (note this depicts only a segment of the mesh). Grey represents TPU and blue represents gel. Nodes have a mobility coefficient $M_\mathbf{n}^\text{TPU,H}$ according to their designation. The edges represent contributions to the spring-like term in \cref{eq:nodalforce}. Edges coloured blue have stiffness coefficient $G^\text{H}$ and those coloured grey $G^\text{TPU}$. Diagonal edges are dashed as a reminder that these are effectively half as stiff as straight edges by \cref{eq:stiffness}.}
  \label{fig:gelTPUboundary}
\end{figure}

Each node on the gel-TPU boundary must be assigned a nodal mobility coefficient and a relevant stiffness value for each `spring' in the spring-like term in \cref{eq:nodalforce} ($G^\text{TPU}$ for springs deemed `TPU' and $G^\text{H}$ for springs deemed `hydrogel'). We choose to assign this as depicted in \cref{fig:gelTPUboundary}.

We implement this algorithm in MATLAB. For time-integration, we use a backward Euler method. We define the residual
\begin{equation}
  R = \mathbf{x}_\mathbf{n}(t)-\mathbf{x}_\mathbf{n}(t-\Delta t)-\Delta t M_\mathbf{n}\mathbf{F}_\mathbf{n},
\end{equation}
and solve the nonlinear equation $R = 0$ using a Newton-Raphson method on each time step. We compute the Jacobian of $R$ analytically by differentiating \cref{eq:nodalforce}.

\section{Results and discussion}
\label{sec:results}
\begin{figure*}[p]
  \centering

  \begin{subfigure}[t]{0.48\textwidth}
    \centering

    \includegraphics[width=0.9\linewidth]{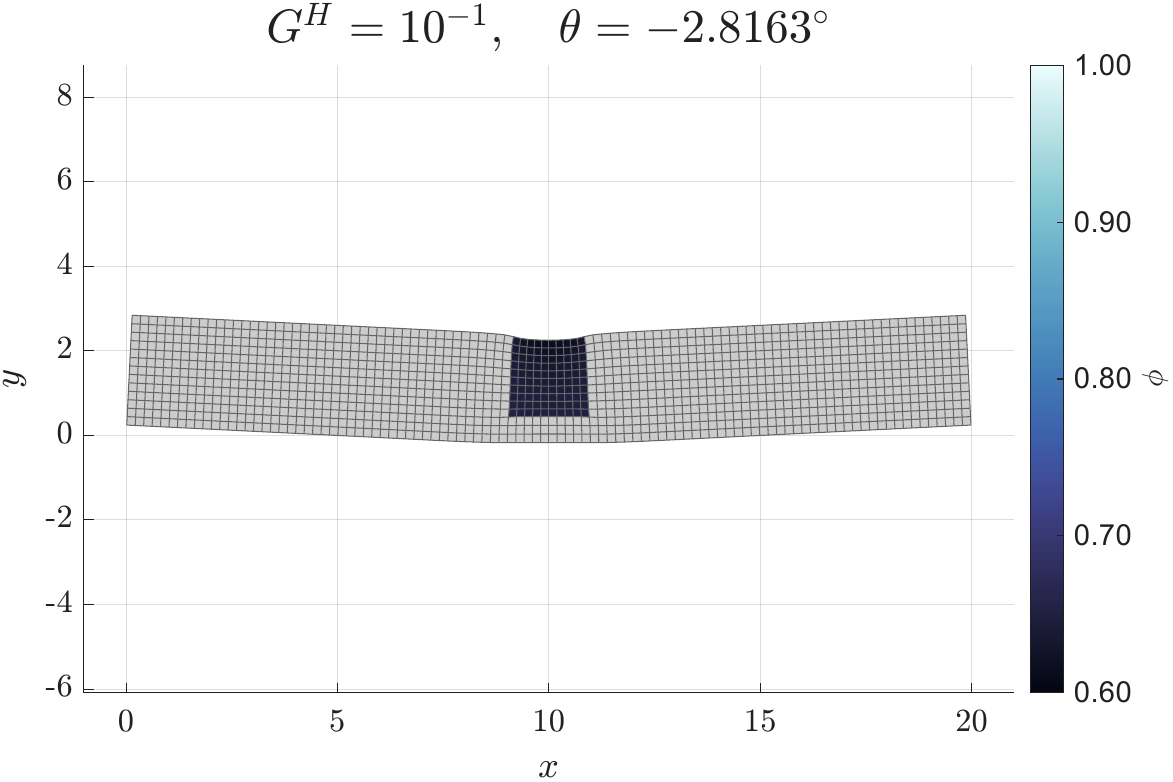}
    \par\vspace{10pt}
    \includegraphics[width=0.9\linewidth]{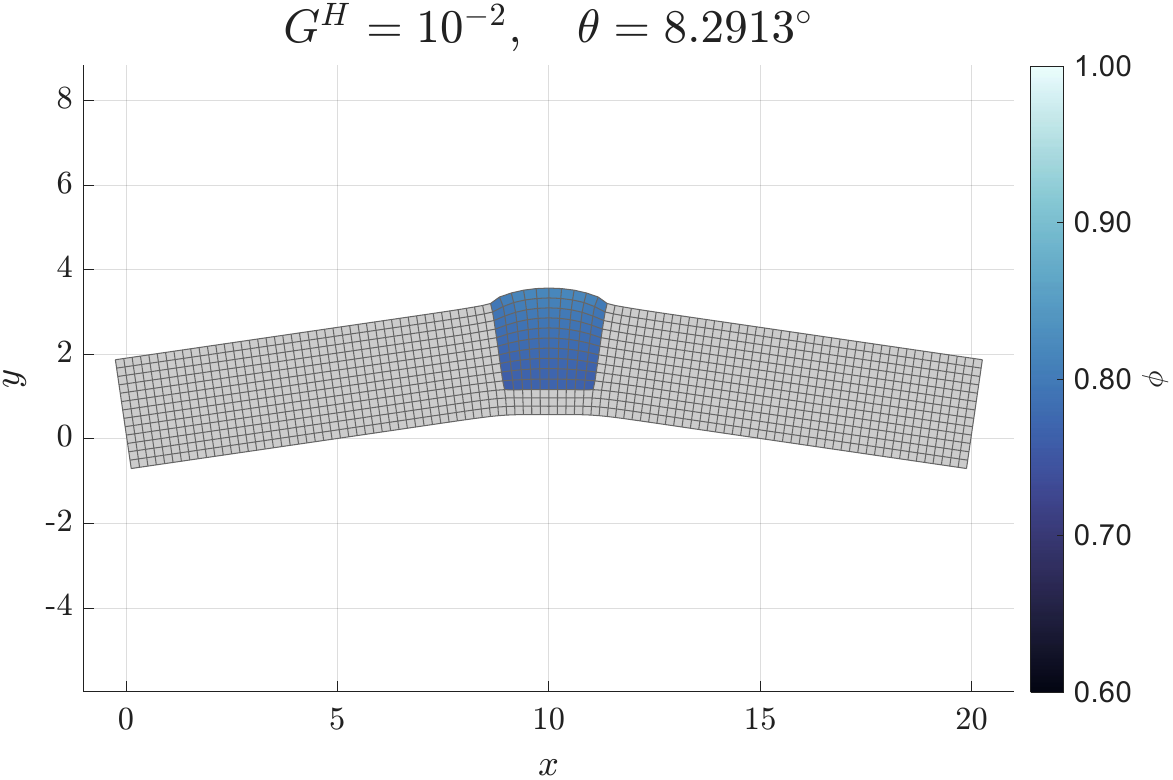}
    \par\vspace{10pt}
    \includegraphics[width=0.9\linewidth]{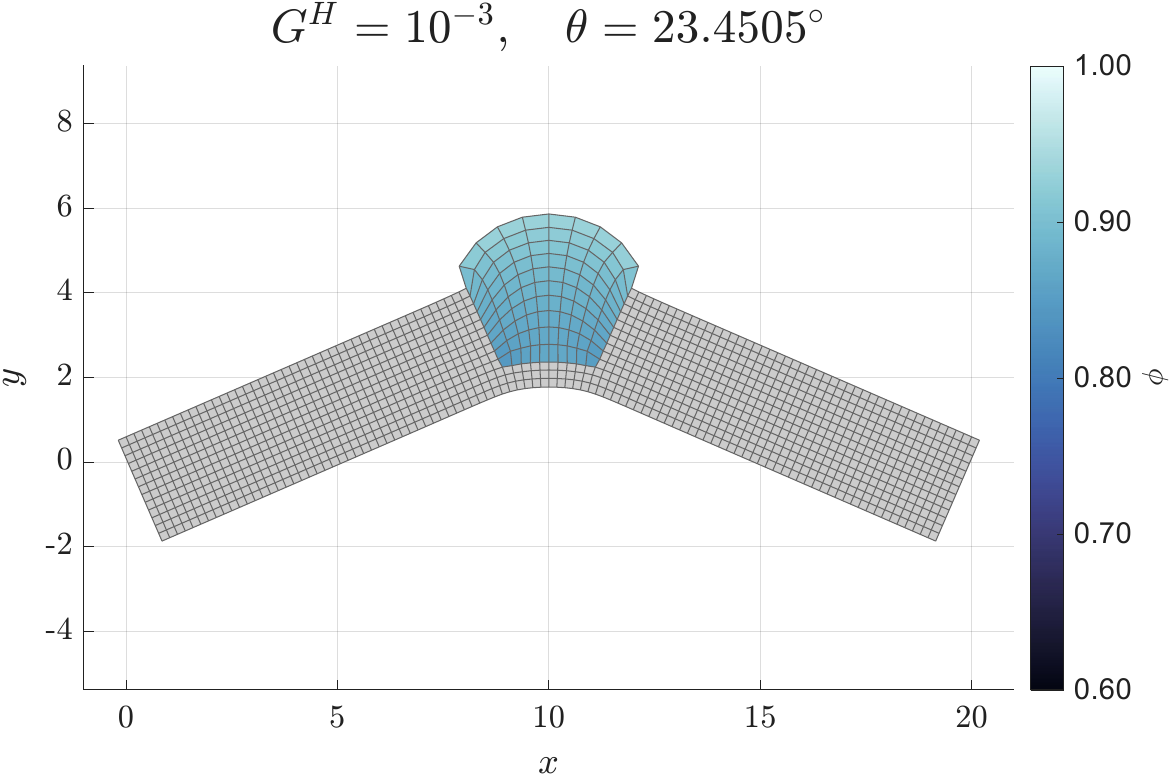}
    \par\vspace{10pt}
    \includegraphics[width=0.9\linewidth]{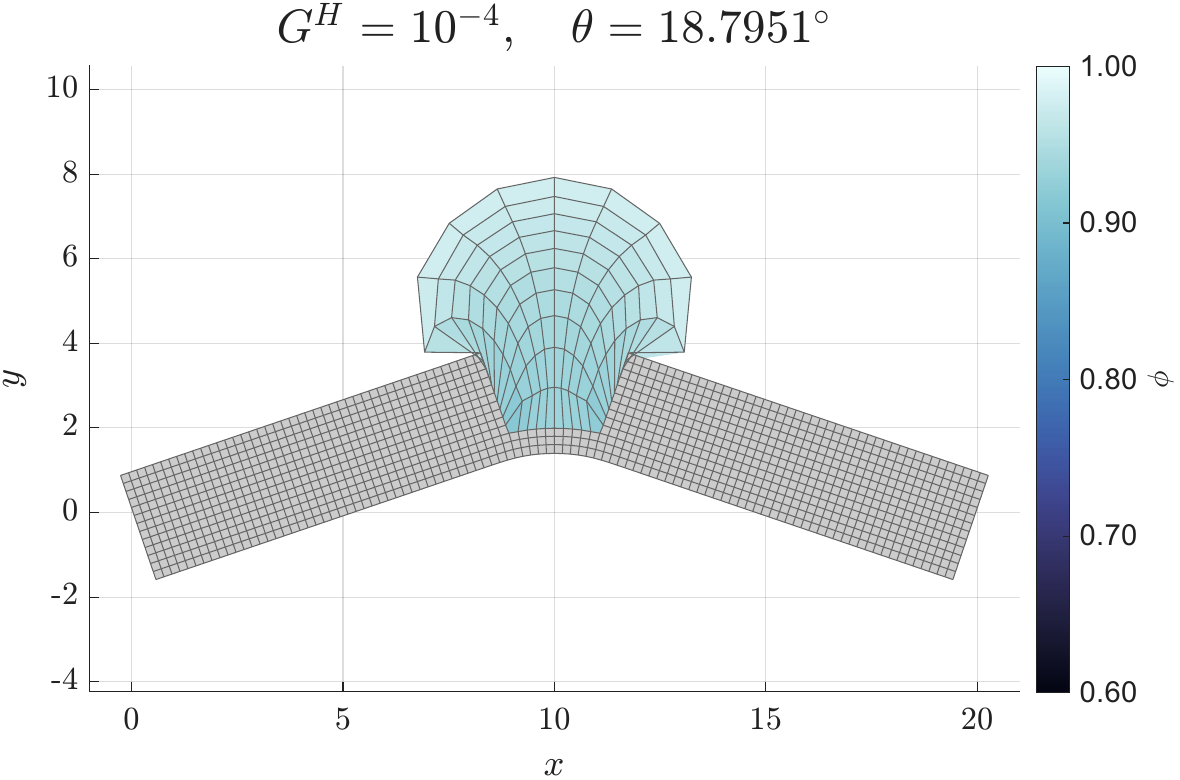}
    \caption{\centering $L^\text{H} = 2$}
    \label{fig:L2}
  \end{subfigure}
  \hfill
  \begin{subfigure}[t]{0.48\textwidth}
    \centering
    \includegraphics[width=0.9\linewidth]{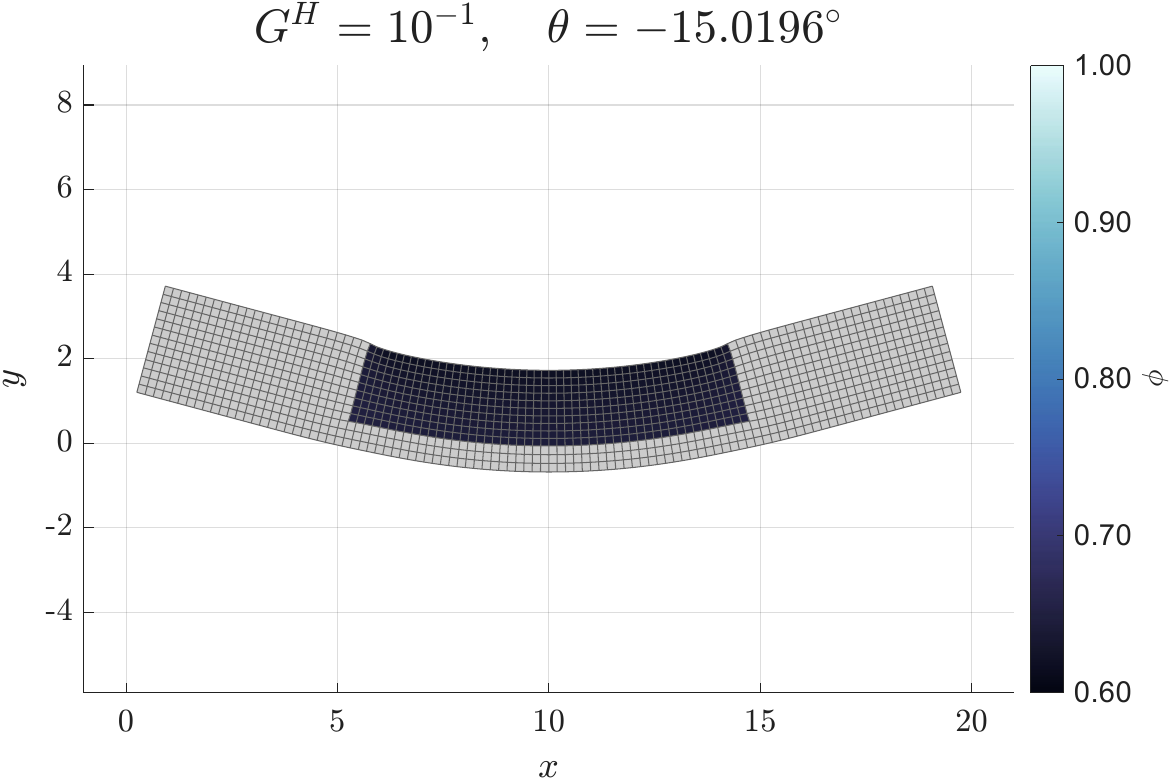}
    \par\vspace{10pt}
    \includegraphics[width=0.9\linewidth]{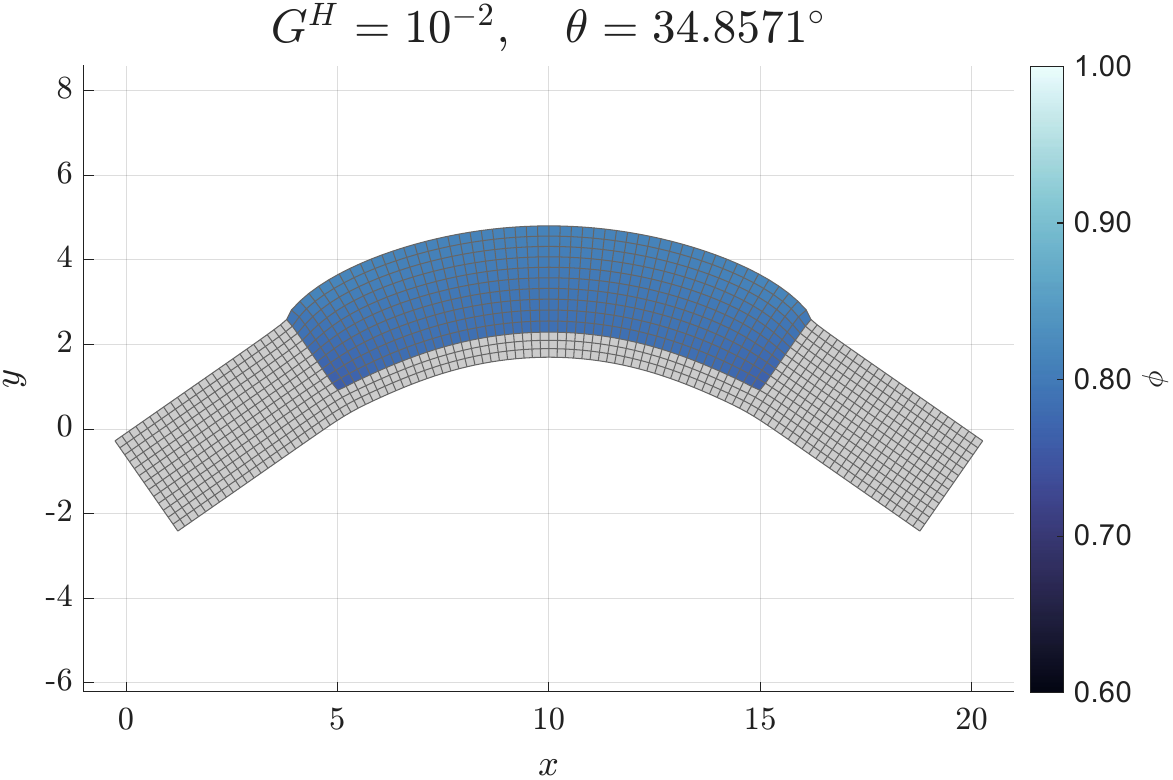}
    \par\vspace{10pt}
    \includegraphics[width=0.9\linewidth]{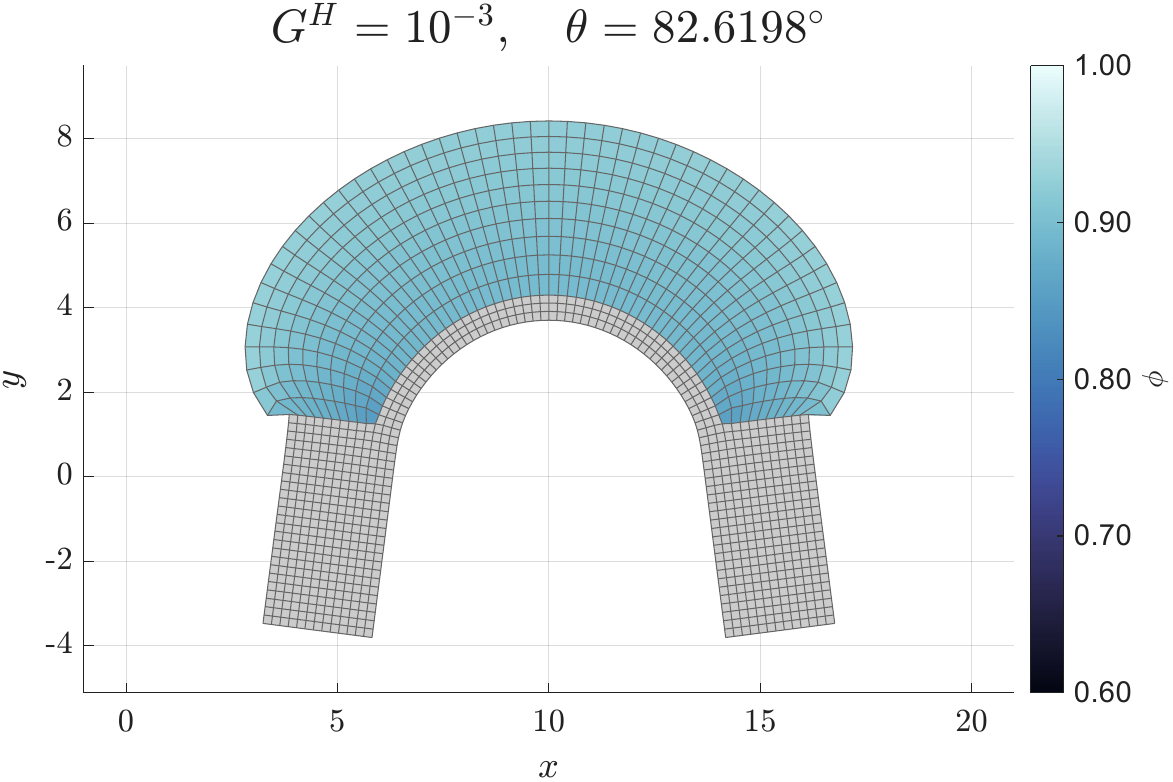}
    \par\vspace{10pt}
    \includegraphics[width=0.9\linewidth]{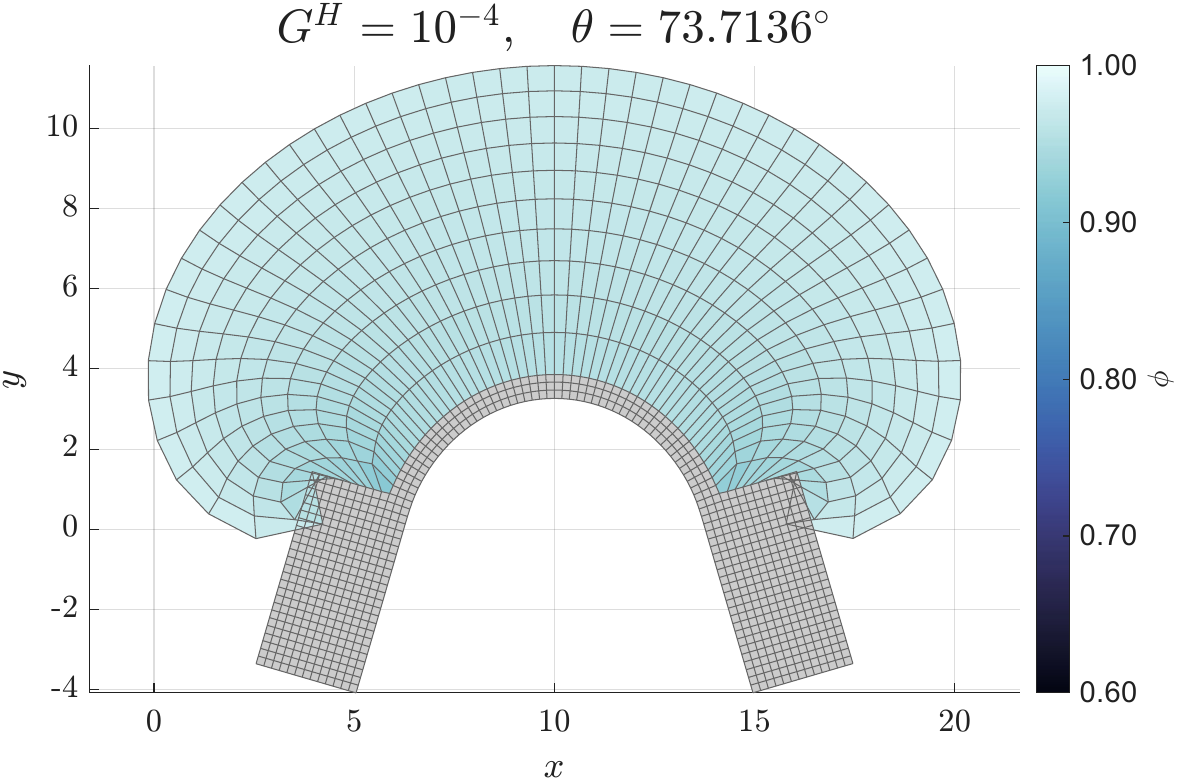}
    \caption{\centering $L^\text{H} = 10$}
    \label{fig:L10}
  \end{subfigure}

  \caption{Hinge endstates for (a) $L^\text{H} = 2$ and (b) $L^\text{H}=10$, with (from top to bottom) $G^\text{H} = 10^{-1}, 10^{-2}, 10^{-3}, 10^{-4}$. Other parameter values are displayed in \cref{tab:params}. Final bend angles $\theta$ are displayed in the titles. For $G^\text{H} = 10^{-4}$, the gel has begun to overlap the TPU at the corners; at the small amount displayed we do not expect this to significantly affect the bending outcome.}
  \label{fig:fullpage_twocol}
\end{figure*}

\begin{figure*}
  \centering
  \begin{subfigure}[t]{0.48\textwidth}
    \centering
    \includegraphics[width=\linewidth]{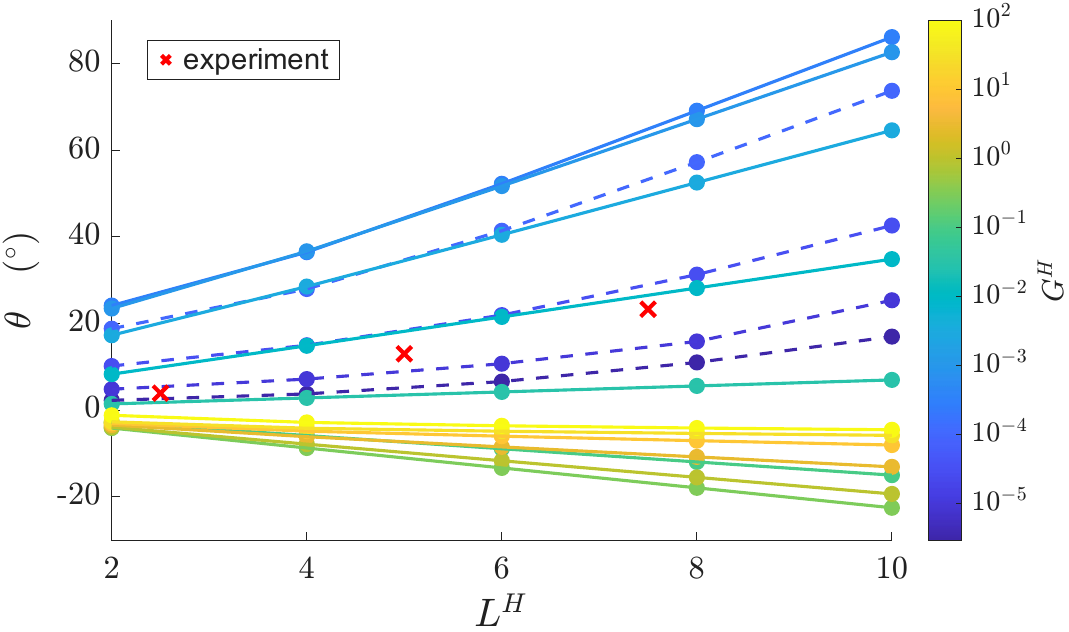}
    \caption{\centering}
    \label{fig:Gscatter}
  \end{subfigure}\hfill
    \begin{subfigure}[t]{0.48\textwidth}
      \centering
    \includegraphics[width=\linewidth]{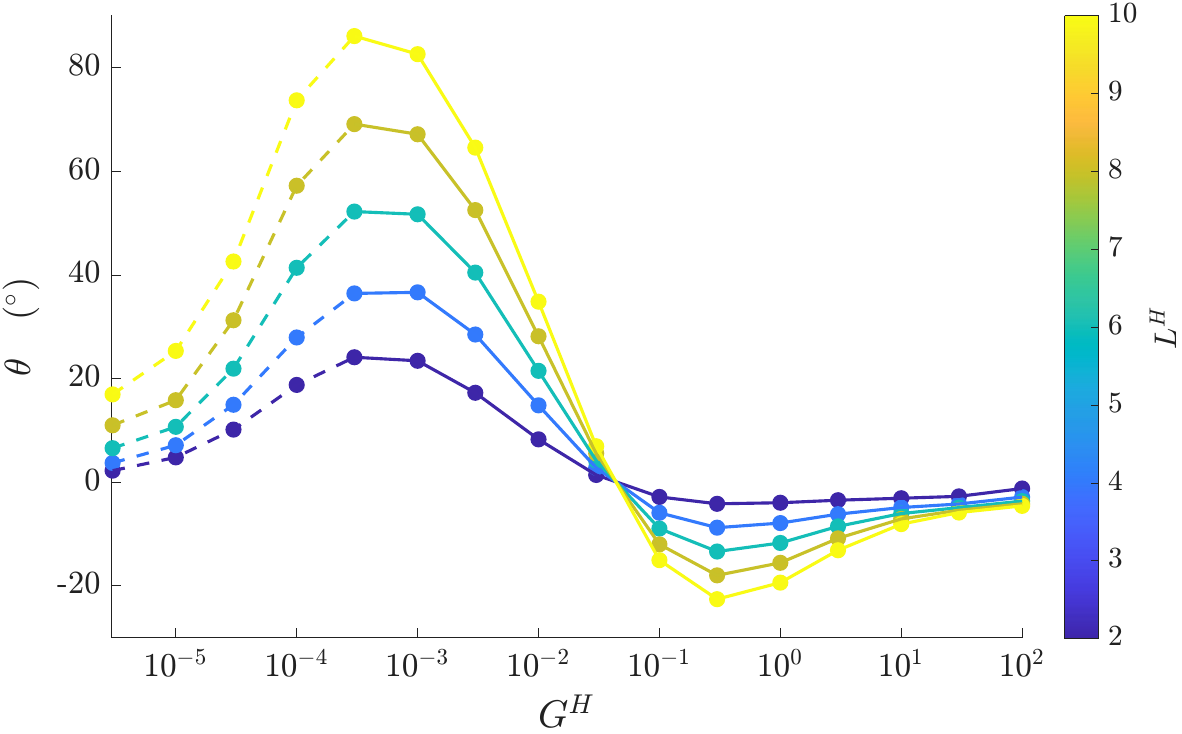}
    \caption{\centering}
    \label{fig:Lscatter}
  \end{subfigure}
  \caption{Scatter plots displaying the effect of $L^\text{H}$ and $G^\text{H}$ on final bend angle $\theta$. Both plots display the same data: in (a) $L^\text{H}$ is on the x-axis and points are coloured by $G^\text{H}$, while in (b) $G^\text{H}$ is on the x-axis and points are coloured by $L^\text{H}$. In (a), we plot the experimental results in red for comparison. Dashed lines indicate $G^\text{H} \leq 10^{-4}$, for which the results have the feature shown in fourth panel of \cref{fig:L10} wherein the gel overlaps the TPU; these should therefore not be taken as predictions for the behaviour of real hinges.}
  \label{fig:scatter}
\end{figure*}

We computed the end states of 80 hinges, with parameters as given in \cref{tab:params}. We varied the length of the gel $L^\text{H}$ and gel stiffness $G^\text{H}$, taking values $L^\text{H} \in [2,10]$ (5 linearly spaced values) and $G^\text{H} \in [3\times 10^{-6},10^{2}]$ (16 logarithmically spaced values). The endstates we obtained for select values are displayed in \cref{fig:fullpage_twocol}, while we plot the final bending angles $\theta$ for each simulation in \cref{fig:scatter}.

\begin{table}
  \begin{tabular}{l l l}
    Parameter & Meaning & Value \\
    \hline
\vspace{-3mm}\\     
    $L^\text{H}$ & length of gel & $2 - 10$ \\
    $L^\text{TPU}$ & length of TPU & 20 \\
    $T^\text{H}$ & thickness of gel & 2 \\
    $T^\text{TPU}$ & thickness of TPU (below gel) & 0.6 \\
    $G^\text{H}$ & shear modulus of gel & $3\times10^{-6} - 10^{2}$ \\
    $G^\text{TPU}$ & shear modulus of TPU & 0.1 \\
    $\lambda$ & bulk modulus of TPU & 10 \\
    $\phi_0$ & initial porosity & 0.7
  \end{tabular}
  \caption{Parameter values for the simulation whose results are shown in \cref{fig:fullpage_twocol,fig:scatter}. Note these are non-dimensional parameters as described in \cref{sec:nond,sec:TPU}. See \cref{fig:hingediagram} for the defintions of $L^\text{H,TPU}, T^\text{H,TPU}$.}
  \label{tab:params}
\end{table}

Firstly, we note that as $G^\text{H}$ decreases, the gel swells more. In particular, for $G^\text{H} = 10^{-4}$ and below, the gel swells sufficiently to overlap the TPU at the corner. The model tracks material position only via edges in the mesh so has no mechanism to prevent this overlap. In theory, this could be accommodated by adding a repulsion potential to nodes and/or springs; however, noting the extent of the swelling displayed in the fourth panel of \cref{fig:L10}, we expect that in reality a gel this soft would fracture before reaching this state. Furthermore, the effect does not occur in optimally bending hinges and gel fracture would only be expected to further reduce the bending from the amount outputted by the model. We therefore opt to include these results as an indication of the overall stiffness trend, while acknowledging that they do not represent sensible predictions of realistic hinge equilibria.

We see from \cref{fig:scatter} that the effect of increasing $L^\text{H}$ is close to linear for all $G^\text{H}$ values sampled (showing possible signs of nonlinearity at larger $L^\text{H}$). The relationship between the $G^\text{H}$ value and the bending angle is more nuanced, in part because of the more nuanced role $G^\text{H}$ plays in the physics: a gel with larger $G^\text{H}$ is not only stiffer but has a lower `desired' equilibrium $\phi$ value. Thus by keeping the initial porosity $\phi_0$ constant at 0.7 in these simulations and varying $G^\text{H}$, the `desired' porosity varies across the simulations. In fact we can see from \cref{fig:Gscatter} that there exists a critical value of $G^\text{H}$ at which the bending angle is approximately zero regardless of $L^\text{H}$, and above this critical value the gel tends to shrink rather than swell. As can be seen in \cref{fig:fullpage_twocol}, this results in bending in the reverse direction. On \cref{fig:Gscatter}, we also plot the experimental results from \cref{fig:hinge_images_experimental} in red for comparison. As can be seen from \cref{fig:hinge_images_experimental}, the experimental hinge begins in a relatively dry state (i.e. the gel does not initially fill the space in the frame left for it) meaning it is not sensible to compare the two directly; however, from the data points we have we can see a similar relation between $L^\text{H}$ and bending angle to that predicted by the model emerging.

In \cref{fig:Lscatter}, we plot the same data as in \cref{fig:Gscatter} (minus the experimental results) but reorganised to display $G^\text{H}$ on the $x$-axis and coloured by $L^\text{H}$. We see that there exists an `optimal' (in the sense of maximising bending angle) for both swelling and deswelling hinges, being near $10^{-3}$ for swelling and near $10^{-1/2}$ for deswelling. The maximum is of greater magnitude than the minimum for this set of parameters; however, this can vary with other parameter values (in particular initial porosity $\phi_0$). Bending seems to approach zero at either extreme. For sufficiently large $G^\text{H}$, it is not energetically viable for a hydrogel with sufficiently large $G^\text{H}$ to swell (increasingly the energy penalty of deforming too great to be countered the favourable change in chemical energy by absorbing/releasing water). Comparing the third and fourth panels of \cref{fig:L2,fig:L10} provides useful intuition for why the bending decreases at the other extreme; although in the fourth panels, with $G^\text{H} = 10^{-4}$, the gel swells more, it is able to do this by `escaping' the frame rather than contributing to bending. This `escaping' eventually leads to a state which is ill-described by the model but it is nontheless illustrative of why a $G^\text{H}$ value too small may be suboptimal for hinge design.

\section{Conclusions}
\label{sec:conc}
In this paper, we have developed a gel lattice spring method for calculating equilibrium states of hydrogel box hinges. The code produces realistic behaviour which agrees qualitatively with experiments (longer hydrogel strips produce more bending).  We used it to examine the effect of two parameters on the eventual bending angle. The results showed that the bending angle $\theta$ depends close to linearly on the length of gel within the hinge $L^\text{H}$, and nonlinearly on the gel stiffness $G^\text{H}$. Depending on the $G^\text{H}$ value, the gel may shrink or swell, and both of these are viable ways to produce bending (in opposite directions). The simulation results indicate there is are two optimum values of $G^\text{H}$, according to whether swelling or deswelling is desired; this is a useful insight which could inform choices of materials for design of DMDs or other actuation devices. The ability for hinges to bend via both swelling and de-swelling indicates the reversibility of this process, a highly beneficial feature for use in biomedical applications as it aids both the insertion and removal processes of DMDs. The simplicity of the dependence of bending angle on $L^\text{H}$ at smaller values offers a naturally useful design heuristic, i.e. doubling gel length can be expected to double the bending angle.

We chose to focus on equilibrium states in this paper; while the structural components required to model dynamics under the assumption that material flux $\mathbf{q} = \mathbf{0}$ are already present in the code, dynamics were outside the scope of the current investigation. While $\mathbf{q} = \mathbf{0}$ is a simplification of the physics, it is unclear how much including the effect of non-zero material flux would impact results, and previous work \citep{yashin2007Theoretical} suggests it produces qualitatively reasonable dynamics. The numerical infrastructure developed in this work is readily adaptible to a range of applications: for example, different hinge geometries, different actuation principles, more complex multi-hinge structures, and even full-scale simulations of origami-inspired devices. The advantage of a lattice spring approach is that it captures interactions with other materials very naturally. Future work could therefore consider simulations of deployed hinges in confined spaces or interactions with other elastic materials.

\section*{Data availability statement}
The code for the gLSM software used in this work can be found at https://github.com/ellen-jolley/Box-hinge-gLSM with DOI 10.5281/zenodo.24139953. The version of the code employed for this study is version 1.0.0.

\section*{Author contributions}
EJ - formal analysis, investigation, methodology, software, visualization, writing (first draft). KP - investigation, methodology, software, writing (first draft). GH - investigation, methodology, resources, visualization, writing (first draft). GC - conceptualization, funding acquisition, investigation, methodology, resources, supervision, writing (reviewing and editing). TDMJ - conceptualization, funding acquisition, investigation, methodology, software, supervision, writing (reviewing and editing).

\section*{Conflicts of interest}
There are no conflicts to declare.

\section*{Acknowledgements}
We thank the Leverhulme Trust which funds EJ, KP and TDMJ. We thank Joseph Webber for helpful comments on a first draft of this manuscript. GH is funded through the EPSRC Centre for Doctoral Training in Topological Design (EP/S02297X/1). 

\section*{AI statement}
The software used in this study was developed by the authors. LLMs were consulted during software development but did not generate the final code. All conceptualization, analysis, and writing involved in the production of this article was undertaken by the authors. The authors take full responsibility for the content of this article.

\bibliographystyle{plainnat}
\bibliography{gLSM_main,gLSM_ellen}

\end{document}